\documentclass[a4paper,11pt]{article}
\usepackage{graphicx} % Required for inserting images
\usepackage[noEucal]{main}
\usepackage{aas_macros}
\usepackage[utf8]{inputenc}

\makeatletter \g@addto@macro\@floatboxreset\centering \makeatother
\usepackage{graphicx}
\usepackage[scr=boondoxo]{mathalfa}
\usepackage[bbgreekl]{mathbbol}
\DeclareSymbolFontAlphabet{\mathbbm}{bbold}
\usepackage[font={small,it}]{caption}

\usepackage{epsfig,latexsym,amsfonts,amsmath,amsthm,amssymb,amsbsy,multirow,slashed,wasysym,textcomp,subfigure,hyperref,wrapfig,datetime,comment,mathtools,cancel,dsfont,diagbox,authblk,braket}

\newcommand{\badat}{\begin{alignedat}}
\newcommand{\eadat}{\end{alignedat}}

\usepackage{color}

\newcommand{\Scal}{\mathcal{S}}

\newcommand{\wb}{\bar{w}}
\newcommand{\xb}{\bar{x}}
\newcommand{\zb}{{\bar{z}}}

\newcommand{\hb}{\bar{h}}

\usepackage[sorting=none,style=numeric-comp,maxbibnames=99]{biblatex}

\numberwithin{equation}{section}

\begin{document}

\begin{titlepage}
  \thispagestyle{empty}
 
  \bigskip\ \bigskip\ \bigskip
  \bigskip\ \bigskip\ \bigskip
  \begin{center}

\baselineskip=12pt {\LARGE \scshape{

Exact Infrared Triangle in Gravity with \\ Long-Range Interactions

}}
  
  \vskip1.5cm 

   \centerline{
Sangmin Choi$^{1,a,b}$,
Ameya Kadhe$^{2,b}$,
Prahar Mitra$^{3,c}$,
and Andrea Puhm$^{4,b}$
   }

\bigskip\bigskip

\centerline{$^{a}$\em {School of Mathematics and Maxwell Institute for Mathematical Sciences,}}\smallskip
\centerline{\em {University of Edinburgh, Edinburgh EH9 3FD, UK}}\smallskip
\centerline{$^{b}$\em {Institute for Theoretical Physics, University of Amsterdam,}}\smallskip
\centerline{\em {PO Box 94485, 1090 GL Amsterdam, The Netherlands}}\smallskip
\centerline{$^{c}$\em {School of Physics \& Astronomy, University of Southampton, Southampton SO17 1BJ, UK}}

\smallskip

\bigskip\bigskip
 \end{center}

\begin{abstract}
Einstein gravity coupled to massless matter exhibits universality in the infrared. In the presence of long-range interactions we show that a novel superrotation charge conservation is equivalent to the massless limit of the classical logarithmic soft graviton theorem, and we propose a formula for the corresponding tail to the displacement memory. These results, which extend earlier work on gravity coupled to massive matter in a qualitatively different way, are exact in the gravitational coupling and reveal that long-range gravitational dressings resum. We also compute higher-order corrections to the known superrotation charge which are exact in the gravitational coupling and would correspond to a 1-loop exact correction to the Cachazo-Strominger subleading soft graviton theorem.

\end{abstract}

\footnotetext[1]{Email: schoi5@ed.ac.uk}
\footnotetext[2]{Email: a.m.kadhe@uva.nl}
\footnotetext[3]{Email: prahar.mitra@soton.ac.uk}
\footnotetext[4]{Email: a.puhm@uva.nl}

\end{titlepage}

%\maketitle
\tableofcontents

\newpage

%%%%%%%%%%%%%%%%%%%%%%%%%%%%%%%%%%%%%%%%%%%%%%%%%%%%%%%%%%%%
\section{Introduction}
\label{sec:intro}
%%%%%%%%%%%%%%%%%%%%%%%%%%%%%%%%%%%%%%%%%%%%%%%%%%%%%%%%%%%%

Gravitational scattering in four dimensions has universal properties at low energies and large distances. This universality is encoded, equivalently, in the soft theorems of scattering amplitudes, in the asymptotic symmetries of asymptotically flat metrics, and in the memory effects on freely falling observers. These three-way equivalences are known as \textit{infrared triangles} \cite{Strominger:2017zoo} (see \cite{Pasterski:2021raf, Raclariu:2021zjz,McLoughlin:2022ljp,Donnay:2023mrd} for reviews). The first instance of such a triangle relates Weinberg's leading soft graviton theorem \cite{Weinberg:1965aa}, BMS supertranslations \cite{Bondi:1962px, Sachs:1962wk, Sachs:1962zza}, and the displacement memory effect \cite{zeldovich, Braginsky:1985vlg, Braginsky:1987kwo, Christodoulou:1991cr}. 
The leading soft graviton theorem is the Ward identity for supertranslations \cite{Strominger:2013jfa, He:2014laa}, and its Fourier transform to retarded time is the displacement memory \cite{Strominger:2014pwa}. Further infrared triangles have since been established: the subleading soft graviton theorem of Cachazo and Strominger \cite{Cachazo:2014fwa} is the Ward identity of superrotations \cite{Barnich:2009se, Barnich:2010eb, Barnich:2011mi, Kapec:2014opa, Campiglia:2014yka, Campiglia:2015yka} and is tied to the spin memory effect \cite{Pasterski:2015tva}. 

While the leading soft theorem is exact to all orders in perturbation theory, the subleading soft theorems receive loop corrections \cite{Bern:2014oka, He:2014bga, Cachazo:2014dia, Bern:2014vva}. These corrections have a physical origin: gravity and electromagnetism are long-range interactions, so the asymptotic trajectories of particles are affected by the gravitational and electromagnetic Coulombic fields generated by all the other particles involved in the scattering process. These interactions affect the soft behaviour of scattering amplitudes in several different but related ways. The tree-level Cachazo-Strominger subleading soft theorem governs the $O(\o^0)$ term in the soft expansion of an amplitude ($\o$ is the energy of the soft graviton). At tree-level, this is subleading to Weinberg's leading soft graviton theorem which appears at $O(\o^{-1})$. It was shown in \cite{Laddha:2018myi, Sahoo:2018lxl} that loop effects in four dimensions introduce terms that are $O(\ln \o)$, which dominate the $O(\o^0)$ term as $\o \to 0$. The coefficient of $\ln\o$ is universal in that it is fixed by the momenta of the hard particles alone, and it is one-loop exact \cite{Sahoo:2018lxl, Saha:2019tub, Sahoo:2021ctw, Krishna:2023fxg} (see \cite{Sen:2024qzb} for a review and \cite{Sahoo:2020ryf, Karan:2025ndk} for higher orders in the soft expansion). Classically, the logarithmic soft graviton theorem fixes the $1/|u|$ fall-off of the gravitational waveform at early and late retarded times \cite{Laddha:2018vbn, Sahoo:2021ctw, Ghosh:2021bam}, the so-called \textit{tail to the displacement memory}, which is the low-energy remnant of the hereditary tail effects of post-Newtonian theory \cite{Blanchet:1992br, Blanchet:2013haa} and has recently been recovered from one-loop scattering amplitudes \cite{Brandhuber:2023hhy, Herderschee:2023fxh, Elkhidir:2023dco, Georgoudis:2023lgf, Caron-Huot:2023vxl, Bini:2023fiz, Georgoudis:2023eke, Alessio:2024onn}. In addition to the $\ln \o$ correction, loop effects also modify the $O(\o^0)$ Cachazo-Strominger soft theorem by an infrared divergent piece \cite{Bern:2014oka, He:2014bga, Cachazo:2014dia, Sahoo:2018lxl, Donnay:2022hkf} which is universal, and a finite regulator-dependent piece \cite{Bern:2014oka, Sahoo:2018lxl, Krishna:2023fxg} (see \cite{Engel:2021ccn} for the analogous one-loop extension of the Low--Burnett--Kroll theorem in QED). As we will see, long-range interactions nevertheless fix a universal, one-loop exact part of the latter.

As with Weinberg's leading soft theorem, the universality of the logarithmic soft theorems calls for a symmetry interpretation. In QED, the loop-corrected subleading soft photon theorem was derived as a Ward identity in \cite{Campiglia:2019wxe} (see \cite{AtulBhatkar:2019vcb} for massless QED coupled to gravity), while in gravity the logarithmic soft graviton theorem was shown to follow from superrotation fluxes through null infinity and their matching across spatial infinity in \cite{Agrawal:2023zea} (see also \cite{Donnay:2022hkf}). In \cite{Choi:2024ygx, Choi:2024ajz, Choi:2024mac} it was shown, for massive matter, that long-range interactions modify the asymptotic symmetry charges themselves: the superrotation (superphaserotation) charges acquire an infrared divergent piece whose coefficient is finite and conserved, and whose conservation law is equivalent to the classical logarithmic soft graviton (photon) theorem. These developments are closely related to the breakdown of peeling and the appearance of polyhomogeneous expansions of the metric near null infinity \cite{Damour:1985cm,Winicour,Chrusciel:1993hx,Kehrberger:2021uvf,Gajic:2022pst,Compere:2023qoa,Geiller:2024ryw,Boschetti:2025tru,Boschetti:2026gfd,Compere:2026jmk}, and to logarithmic asymptotic symmetries at spatial and null infinity \cite{Campiglia:2018dyi, Fuentealba:2022xsz, Fuentealba:2024lll, Fuentealba:2025ekj, Girelli:2026gbr, Fuentealba:2026txc}.

In the approach of \cite{Choi:2024ygx, Choi:2024ajz, Choi:2024mac}, only massive matter in gravity has been treated so far (the derivation of \cite{Agrawal:2023zea} applies to massless particles as well, but the logarithms there are obtained from the quantum Goldstone mode rather than from the dressing of the asymptotic fields). The case of massless matter is both more subtle and qualitatively different: there are no contributions from timelike infinity, so the long-range dressing of the hard particles must be captured entirely at null infinity, where it modifies the free-field fall-offs of both the matter and the radiation fields by overleading logarithmic terms that have to be resummed. In \cite{Choi:2025mzg}, this analysis was carried out for massless scalar QED where it was established that the hard logarithmic superphaserotation charge vanishes identically to all orders in the electromagnetic coupling and that there is no tail to the velocity kick memory. This is in agreement with the vanishing of the classical logarithmic soft photon theorem for massless charged particles. Thus all corners of the classical logarithmic infrared triangle of massless scalar QED exactly vanish. The same is not true in gravity. In this case, the massless limit of the classical logarithmic soft graviton theorem is known to be finite and non-vanishing \cite{Sahoo:2018lxl}. In this paper, we follow the approach of \cite{Choi:2024ygx, Choi:2024ajz, Choi:2024mac} and extend the infrared triangle, established there for massive matter in gravity and QED and in \cite{Choi:2025mzg} for massless matter in QED, to Einstein gravity coupled to massless matter.

For concreteness, we consider a massless scalar minimally coupled to Einstein gravity. We study the theory perturbatively in the gravitational coupling $\k = \sqrt{32\pi G_N}$. Working in retarded coordinates $(u,r,{\hat x})$ (${\hat x}$ is a three-dimensional unit vector labelled by a point on the celestial sphere) and in de Donder gauge we will show that long-range interactions generate a $u$-independent $1/r$ term in $h_{rr}$, whose coefficient we denote by $\SP ({\hat x})$. It is given by
\begin{equation}
\begin{split}
\SP({\hat x}) =  \frac{\k}{8\pi} ( E - \vec{P} \cdot {\hat x} ) ,
\end{split}
\end{equation}
where $(E,\vec{P})$ is the total energy-momentum passing through future null infinity $\CI^+$. This Coulombic mode generates an infinite tower of overleading logarithms in both the graviton field and the massless scalar field. We argue that the tower resums, to all orders in $\k$, into a shift of the retarded time argument of the outgoing radiative data $\varphi^\text{out}$ and $h^\text{out}_{AB}$. More precisely, we will see that the scalar field and the trace-free part of the transverse metric perturbation behave near $\CI^+$ as\footnote{Here, $L$ is the length scale that characterises the size of the scattering region \cite{Sahoo:2018lxl}. This length scale is distinct from the infrared cutoff length scale $R$. The limit we are interested in is $R/L \gg 1$.}
\begin{equation}
\begin{split}
\label{ushift_intro}
\varphi(u,r,{\hat x}) = \frac{1}{r} \varphi^\text{out} \left( u-\frac{\k}{2} \SP ({\hat x}) \ln\frac{r}{L} ,  {\hat x} \right) + o(r^{-1})  , \\
h_{\{AB\}} (u,r,{\hat x}) =  r h^\text{out}_{AB} \left( u-\frac{\k}{2} \SP ({\hat x}) \ln\frac{r}{L} ,  {\hat x} \right) + o(r) .
\end{split}
\end{equation}
The physical interpretation of this shift will be discussed in \cite{Choi:2026xxx} (see also \cite{Fuentealba:2026txc} for related recent work). The shift \eqref{ushift_intro} renders the superrotation charges logarithmically divergent. These divergences are regulated with a large-distance cutoff $r=R$ and the charges split into a logarithmically divergent term (proportional to $\ln \frac{R}{L}$) and a finite part. The coefficient of $\ln \frac{R}{L}$ is conserved and the associated Ward identity is precisely the massless limit of the classical logarithmic soft graviton theorem of \cite{Sahoo:2018lxl}. The finite part of the charge reproduces the superrotation charges of \cite{Kapec:2014opa}, and hence the Cachazo--Strominger soft theorem, supplemented by a $\k$-exact Coulombic correction which we interpret as a one-loop exact correction to the subleading soft graviton theorem.

Finally, solving Einstein's equations with retarded boundary conditions, we obtain the displacement memory sourced by a massless scalar field, the field-theoretic version of the memory formula for massless particles \cite{Braginsky:1987kwo, Bieri:2013ada, Tolish:2014bka}, and propose a formula for its tail, the $1/|u|$ term in the late-time shear, in terms of the interacting matter and graviton stress tensors; in contrast to massless QED \cite{Choi:2025mzg}, the tail does not vanish.

This paper is organised as follows. In Section~\ref{sec:asymp_data}, we discuss the perturbative expansion of the metric and scalar fields in the full interacting theory. In Section~\ref{sec:large_diffeos_charge}, we introduce the superrotation vector fields compatible with these asymptotics and the de Donder gauge conditions, construct the associated Noether charges in the interacting theory and extract the regulated hard and soft charges. 
In Section~\ref{sec:softthem_derivation}, we derive the logarithmic soft-graviton theorem from the conservation of the log superrotation charge and discuss corrections to the Cachazo-Strominger subleading soft-graviton theorem due to long-range interactions. Finally, in Section~\ref{sec:memory}, we discuss the displacement memory and its tail. In Appendix \ref{app:retarded_sol}, we determine the large $r$ expansion of the linearised perturbation with retarded boundary conditions.

%%%%%%%%%%%%%%%%%%%%%%%%%%%%%%%%%%%%%%%%%%%%%%%%%%%%%%%%%%%%
\section{Asymptotic data of metric and matter fields}
\label{sec:asymp_data}
%%%%%%%%%%%%%%%%%%%%%%%%%%%%%%%%%%%%%%%%%%%%%%%%%%%%%%%%%%%%

Though our results are general and hold for generic gravitational theories (details are presented in the follow-up work \cite{Choi:2026xxx}), we consider for concreteness a massless scalar $\varphi$ minimally coupled to Einstein gravity. This is described by the action
\begin{equation}
\label{massless_action}
S[g,\varphi]=\int_{\CM} \dt^4 x \sqrt{-g}\left( \frac{2}{\k^2} \CR-\frac{1}{2}g^{\mu\nu}\p_\mu \varphi\p_\nu \varphi \right) , \qquad \k=\sqrt{32\pi G_{N}} . 
\end{equation}
The equations of motion that follow from this are
\begin{align}
\label{eom_grav}
\CR_{\mu\nu} - \frac{1}{2} g_{\mu\nu} \CR &= \frac{\k^2}{4} T^\varphi_{\mu\nu}  , \\
\label{eom_scalar}
\frac{1}{\sqrt{-g}} \p_\mu \left( \sqrt{-g} g^{\mu\nu} \p_\nu \varphi \right) &= 0 ,
\end{align}
where
\begin{align}
\label{matter_st}
T^\varphi_{\mu\nu} = \p_\mu \varphi\p_\nu \varphi- \frac{1}{2} g_{\mu\nu}  g^{\a\b} \p_\a\varphi\p_\b\varphi. 
\end{align}
The action and equations of motion \eqref{massless_action}--\eqref{eom_scalar} are all invariant under diffeomorphisms which infinitesimally act on the fields as
\begin{equation}
\begin{split}
\label{diffeos_gdd}
\d_\xi g_{\mu\nu} = \xi^\a \p_\a g_{\mu\nu} + g_{\mu\a} \p_\nu \xi^\a + g_{\a\nu} \p_\mu \xi^\a , \qquad \d_\xi \varphi = \xi^\a \p_\a \varphi .
\end{split}
\end{equation}
Small diffeomorphisms (corresponding to vectors $\xi^\mu$ that vanish on $i^0$) represent redundancies in our description of the theory. These redundancies are removed by imposing a gauge condition that we will describe shortly.

We work perturbatively around Minkowski spacetime where the metric in Cartesian coordinates $X^\a$ is
\begin{equation}
\begin{split}
\dt s^2 = \eta_{\a\b} \dt X^\a \dt X^\b = - (\dt X^0)^2 + \dt \vec{X} \cdot \dt \vec{X} . 
\end{split}
\end{equation}
To describe the asymptotics near $\CI^+$, we will work in retarded coordinates $(u,r,x^A)$ which are related to Cartesian coordinates via
\begin{equation}
\begin{split}
\label{qt_def}
X^\a = u t^\a  + r q^\a({\hat x}) ,\qquad t^\a = (1,0,0,0), \qquad q^\a({\hat x}) = (1,{\hat x}). 
\end{split}
\end{equation}
Here, we represent the point on $S^2$ with angular coordinates $x^A$ by the unit vector ${\hat x} \in \mrr^3$ (${\hat x} \cdot {\hat x} = 1$). The metric in these coordinates is 
\begin{equation}
\label{mink_metric_retarded}
\dt s^2 = \eta_{\mu\nu} \dt x^\mu \dt x^\nu = -\dt u^2-2\dt u\dt r+r^2 \g_{AB} \dt x^A \dt x^B . 
\end{equation}
$\g_{AB}({\hat x})$ is the unit round metric on $S^2$. We denote by $\n_\mu$ the covariant derivative compatible with $\eta_{\mu\nu}$ and by $D_A$ the covariant derivative compatible with $\g_{AB}$. In retarded coordinates, future null infinity $\CI^+$ is located at $r\to\infty$ keeping $(u,{\hat x})$ fixed. While the calculations explicitly presented in this paper are performed near $\CI^+$, analogous formulas can be derived near $\CI^-$ as well. 

The metric perturbation $h_{\mu\nu}$ is defined by
\begin{equation}
\label{hpert_def}
g_{\mu\nu}=\eta_{\mu\nu}+\k  h_{\mu\nu}.
\end{equation}

The diffeomorphisms \eqref{diffeos_gdd} act on the metric perturbation via
\begin{equation}
\begin{split}
\label{h_diff_transform}
\d_\xi h_{\mu\nu} = \frac{1}{\k} ( \n_\mu \xi_\nu + \n_\nu \xi_\mu )  + \CL_\xi h_{\mu\nu} . 
\end{split}
\end{equation}
The gauge redundancies described earlier are removed by working in de Donder or harmonic gauge,\footnote{The de Donder gauge condition \eqref{harmonic_gauge} does not completely remove all the gauge redundancies in the theory, but these residual small diffeomorphisms will not play a role in this paper.}
\begin{equation}
\label{harmonic_gauge}
\n^\nu \hb_{\mu\nu} = 0 , \qquad   \hb_{\mu\nu} \equiv h_{\mu\nu}-\frac{1}{2}\eta_{\mu\nu} h , \qquad h=\eta^{\mu\nu}h_{\mu\nu} .
\end{equation}
Here ${\bar h}_{\mu\nu}$ is the trace-reversed perturbation. Greek indices are raised and lowered with respect to $\eta_{\mu\nu}$ and uppercase Latin indices are raised and lowered with respect to $\g_{AB}$.

In de Donder gauge and to leading order in $\k$, the scalar equation of motion \eqref{eom_scalar} takes the form
\begin{equation}
\label{scalareq_leading}
( \eta^{\mu\nu} - \k h^{\mu\nu} ) \n_\mu \n_\nu  \varphi =  O(\k^2)  .
\end{equation}
In this gauge, Einstein’s equation can be written in the form
\begin{equation}
\label{Einstein_leading}
\n^2 \hb_{\mu\nu}=-\frac{\k}{2} T_{\mu\nu} , \qquad T_{\mu\nu} \equiv T^{\varphi}_{\mu\nu}+T^{\text{H}}_{\mu\nu} , 
\end{equation}
where
\begin{align}
\label{st_scalar_leading}
T^{\varphi}_{\mu\nu} &= \p_\mu \varphi\p_\nu \varphi-\frac{1}{2} \eta_{\mu\nu} \eta^{\rho\s} \p_\rho \varphi \p_\s \varphi-\frac{\k}{2}\left( h_{\mu\nu}\eta^{\rho\s}-\eta_{\mu\nu} h^{\rho\s}\right)\p_\rho \varphi\p_\s \varphi + O(\k^2 h^2) , \\
T^{\text{H}}_{\mu\nu} &=2h^{\rho\s}\left[2\n_\rho \n_{(\mu}h_{\nu)\s}-\n_\mu \n_\nu h_{\rho\s}-\n_\rho \n_\s h_{\mu\nu}\right] - h_{\mu\nu}\n^2h - 4 \n_{[\rho} h_{\s]\mu} \n^{[\rho} h^{\s]}{}_\nu \nonumber 
\\
\label{st_grav_leading}
&\qquad - \n_\mu h_{\rho\s}\n_\nu h^{\rho\s}+\eta_{\mu\nu} \left( 2 h^{\rho\s} \n^2 h_{\rho\s} + 2 \n^{[\rho} h^{\s]\tau} \n_{[\rho} h_{\s]\tau}  +  \frac{1}{2} \n^\tau h^{\rho\s}  \n_\tau h_{\rho\s}  \right) + O(\k h^3) .
\end{align}
Here, we show explicitly the terms in the effective stress tensor that we will need later in the paper.

In the following subsections, we will solve \eqref{eom_scalar} and \eqref{Einstein_leading} in an asymptotic expansion near $\CI^+$ (i.e. near $r = \infty$) to obtain the asymptotic behaviour of $\varphi$ and $h_{\mu\nu}$ near $\CI^+$. We will focus on the behaviour near future null infinity but a similar analysis can be carried out at past null infinity.

%%%%%%%%%%%%%%%%%%%%%%%%%%%%%%%%%%%%%%%%
\subsection{Free fields}
\label{sec:free_fields}
%%%%%%%%%%%%%%%%%%%%%%%%%%%%%%%%%%%%%%%%

To solve the equations of motion, we expand the scalar field and the metric perturbation in powers of $\k$,
\begin{equation}
\begin{split}
\label{pert_exp}
\varphi = \sum_{n=0}^\infty \k^n \varphi_{(n)}   , \qquad h_{\mu\nu}  =  \sum_{n=0}^\infty \k^n h_{(n)\mu\nu}   . 
\end{split}
\end{equation}
Substituting this into \eqref{scalareq_leading} and \eqref{Einstein_leading}, we find at leading order that
\begin{equation}
\begin{split}
\n^2 \varphi_\0 = 0 , \qquad \n^2 {\bar h}_{\0\mu\nu} = 0 . 
\end{split}
\end{equation}
The free fields satisfy the wave equation in Minkowski spacetime. We therefore write the solutions as mode expansions in Fourier space and use these to extract their asymptotic behaviour.

%%%%%%%%%%%%%%%%%%%%%%%%%%%%%%%%%%%%%%%%
\subsubsection{Graviton field}
\label{sec:radiation}
%%%%%%%%%%%%%%%%%%%%%%%%%%%%%%%%%%%%%%%%
The mode expansion for the free gravitational perturbation field $h_{\0\mu\nu}$ is
\begin{equation}
\label{grav_mode_exp}
h_{\0\a\b}(X)=\int\frac{\dt ^{3}k}{(2\pi)^{3}(2|\vec{k}|)}\left[a_{\a\b}(\vec{k})e^{ik\cdot X}+a_{\a\b}^\dagger (\vec{k}) e^{-ik\cdot X}\right], \qquad a_{\a\b}(\vec{k}) = \sum_{\l=\pm} \ve^{\l*}_{\a\b} (\vec{k}) a_\l(\vec{k}) ,
\end{equation}
where $k^{0}=|\vec{k}|$, $k\cdot X=- |\vec{k}| X^0 +\vec{k}\cdot \vec{X}$. The large $r$ expansion of this can be evaluated using a saddle point analysis (see e.g. \cite{He:2014laa}) and we find
\begin{equation}
h_{\0\a\b}(u,r,{\hat x})=-\frac{i}{8\pi^2r}\int_{0}^{\infty}\dt \o \left[a_{\a\b}(\o ,\hat{x})e^{-i\o u}-a^{\dagger}_{\a\b}(\o ,\hat{x}) e^{i\o u}\right] + O(1/r^2).
\end{equation}
Here, the saddle point analysis localises the integral over ${\hat k}$ to ${\hat x}$. The integral over $\o=|\vec{k}|$ still remains after this localisation. 

In four dimensions, gravitons have two radiative degrees of freedom, captured by two polarisations. We work in the helicity basis where the polarisations are $\pm$ and the associated polarisation tensors are
\begin{equation}\label{graviton polarisation}
\ve^\l_{\a\b}(\vec{k})= \ve^\l_\a(\vec{k}) \ve^\l_\b(\vec{k}) , \qquad \ve^+_\a(\vec{k}) =\frac{1}{\sqrt2}(-\zb,1,-i,-\zb) , \qquad \ve^{-}_\a(\vec{k}) = \frac{1}{\sqrt2}(-z,1,i,-z),
\end{equation}
where $z=e^{i\phi}\tan\frac{\t}2$ is the complex stereographic coordinate for the unit 2-sphere. With this choice of polarisation the large $r$  behaviour of the components of $h_{\0\mu\nu}$ in retarded coordinates is
\begin{equation}
\begin{split}
\label{hfree}
h_{\0uu}&=O(r^{-1}), \qquad h_{\0ur}=O(r^{-2}), \qquad h_{\0uA}=O(r^{0}), \\
h_{\0rr} &= O(r^{-3}) , \qquad h_{\0rA}=O(r^{-1}), \qquad h_{\0AB}=O(r).
\end{split}
\end{equation}
Using \eqref{st_grav_leading}, these fall-offs give
\begin{equation}
\label{grav_st_large_r}
\badat2
T_{\0uu}^h=O(r^{-2}), \qquad &T_{\0ur}^h=O(r^{-4}), \qquad T_{\0uA}^h=O(r^{-2}), \\
T_{\0rr}^h=O(r^{-4}), \qquad & T_{\0rA}^h=O(r^{-3}),\qquad T_{\0AB}^h=O(r^{-1}).
\eadat
\end{equation}
The radiative part of the graviton field at $O(\k^0)$ is the $O(r)$ coefficient of $h_{AB}$ and this is given by \begin{equation}
\begin{split}
\label{grav_rad_def}
\overset{-1,0}{h_{\0AB}}(u,\hat{x}) =  - \frac{i}{8\pi^2} \int_{0}^{\infty}\dt \o \left[  a_{AB} (\o ,\hat{x})e^{-i\o u}-a_{AB}^\dagger (\o ,\hat{x})e^{i\o u}\right] ,
\end{split}
\end{equation}
where 
\begin{equation}
\begin{split}
a_{AB}(\o,{\hat x}) = \p_A q^\a({\hat x}) \p_B q^\b({\hat x}) a_{\a\b}(\o,{\hat x}) . 
\end{split}
\end{equation}
In \eqref{grav_rad_def} and in the rest of this paper, we use the notation $\overset{m,n}f$ to denote the coefficient of $r^{-m} (\ln\frac{r}{L})^n$ of the function $f$ at large $r$.

%%%%%%%%%%%%%%%%%%%%%%%%%%%%%%%%%%%%%%%%
\subsubsection{Scalar field}
\label{sec:free_scalar}
%%%%%%%%%%%%%%%%%%%%%%%%%%%%%%%%%%%%%%%%
The mode expansion for the massless free scalar field is 
\begin{equation}
\varphi_\0(X)=\int_{\mrr^3} \frac{\dt ^{3}k}{(2\pi)^{3}(2|\vec{k}|)} \left( a(\vec{k})e^{ik\cdot X}+a^{\dagger}(\vec{k})e^{-ik\cdot X}\right) .
\end{equation}
The saddle point analysis implies that near $\CI^+$, we have
\begin{equation}
\label{scalar_saddle_point}
\varphi_\0(u,r,{\hat x})=-\frac{i}{8\pi^2r}\int_{0}^{\infty}\dt \o \left[a(\o ,\hat{x})e^{-i\o u}-a^{\dagger}(\o ,\hat{x})e^{i\o u}\right]+O(1/r^2).
\end{equation}
The large $r$ behaviour of the free scalar is therefore given by
\begin{equation}
\label{free_scalar_large_r}
\varphi_\0(u,r,\hat{x}) ~\overset{r\to\infty}{=}~ \frac{1}{r} \overset{1,0}{\varphi_\0}(u,\hat{x})+O(1/r^2),
\end{equation}
where 
\begin{equation}
\begin{split}
\label{phi_rad_def}
\overset{1,0}{\varphi_\0}(u,\hat{x}) =  -\frac{i}{8\pi^2}\int_{0}^{\infty}\dt \o \left[a(\o ,\hat{x})e^{-i\o u}-a^{\dagger}(\o ,\hat{x})e^{i\o u}\right]  . 
\end{split}
\end{equation}
The matter stress tensor at $O(\k^0)$ is 
\begin{equation}
\begin{split}
T^{\varphi}_{\0\mu\nu} &= \p_\mu \varphi_\0 \p_\nu \varphi_\0 -\frac{1}{2} \eta_{\mu\nu} \nabla^\rho \varphi_\0 \nabla_\rho \varphi_\0 . 
\end{split}
\end{equation}
Then, using \eqref{free_scalar_large_r}, we find the large $r$ asymptotics of the free matter stress tensor to be 
\begin{equation}
\label{scalar_st_large_r}
\badat2
T_{\0uu}^\varphi &= O(r^{-2}), \qquad  T_{\0ur}^\varphi = O(r^{-4}), \qquad T_{\0uA}^\varphi = O(r^{-2}),
\\
T_{\0rr}^\varphi &= O(r^{-4}), \qquad T_{\0rA}^\varphi = O(r^{-3}), \qquad T_{\0AB}^\varphi = O(r^{-1}).
\eadat
\end{equation}
Note that these are the same fall-offs as those of the graviton stress tensor \eqref{grav_st_large_r}.

%%%%%%%%%%%%%%%%%%%%%%%%%%%%%%
\subsection{Interacting fields}
\label{sec:interacting_fields}
%%%%%%%%%%%%%%%%%%%%%%%%%%%%%%

In the previous section, we evaluated the large $r$ fall-offs for the fields at $O(\k^0)$ where both the massless scalar and the graviton are free. We now discuss how long-range interactions in four dimensions modify the free asymptotics \eqref{hfree} and \eqref{free_scalar_large_r}. We first discuss the modified asymptotics to $O(\k^2)$, and then conjecture an extension to all orders in $\k$. In our follow-up work \cite{Choi:2026xxx}, we show that our conjecture holds non-perturbatively in $\k$.

%%%%%%%%%%%%%%%%%%%%%%%%%%%%%%
\subsubsection{Graviton field}
%%%%%%%%%%%%%%%%%%%%%%%%%%%%%%

\paragraph{Solution at $O(\k)$:} At this order, the equation of motion \eqref{Einstein_leading} for the graviton field simplifies to
\begin{equation}
\label{first_order_eq_grav}
\n^2 \hb_{\1\mu\nu}=-\frac{1}{2} T_{\0\mu\nu} , \qquad T_{\0\mu\nu}  \equiv T^{\varphi}_{\0\mu\nu}+T^{\text{H}}_{\0\mu\nu} . 
\end{equation}
With retarded boundary conditions, the solution is\footnote{We also recall that the de Donder gauge conditions do not uniquely fix the gauge redundancies. In writing the solution \eqref{h_ret_sol}, we have made implicit choices that gauge fix the residual small diffeomorphisms in de Donder gauge.}
\begin{equation}
\begin{split}
\label{h_ret_sol}
{\bar h}_{\1\a\b}(X)  = \frac{1}{4\pi} \int \dt^4 X' \d((X-X')^2) \t(X^0-X'^0) T_{\0\a\b}(X') , 
\end{split}
\end{equation}
Note that since we are using the retarded propagator to solve for the interacting solution, the creation and annihilation operators appearing in \eqref{grav_mode_exp} are the $in$ operators. They live on $\CI^-$ and create or annihilate incoming gravitons in a scattering amplitude. Although evaluated here near $\CI^+$, $h_{\0\a\b}$ is the free solution determined by incoming data on $\CI^-$. The full outgoing radiative data include both this freely propagated contribution and the corrections due to interactions. From here on, we will add a superscript ``in'' to the creation and annihilation operators to clarify this so that
\begin{equation}
\begin{split}
\label{grav_rad_def_in}
\overset{-1,0}{h_{\0AB}}(u,\hat{x}) &=  - \frac{i}{8\pi^2} \int_{0}^{\infty}\dt \o \left[  a^\text{in}_{AB} (\o ,\hat{x}) e^{-i\o u}-a_{AB}^{\text{in}\dagger} (\o ,\hat{x})e^{i\o u}\right] \equiv h_{AB}^\text{in}(u,{\hat x}) . \\
\end{split}
\end{equation}
We can use \eqref{h_ret_sol} to evaluate the asymptotic expansion of the metric perturbation. This is done in Appendix \ref{app:retarded_sol}. The leading $1/r$ coefficient of $h_{\1\a\b}$ in Cartesian coordinates is given by \eqref{app:hab_final}. Using this, we obtain the following asymptotic behaviour of the metric components in retarded coordinates:
\begin{equation}
\begin{split}
\label{hdd_falloffs_1}
h_{\1uu} &= O ( r^{-1} \ln(r/L)) , \qquad h_{\1ur} = O(r^{-1}) , \qquad h_{\1uA} = O(r^0) ,\\
h_{\1rr} &= O(r^{-1}) , \qquad \qquad \quad\,\, h_{\1rA} = O(r^0) , \qquad\,\, h_{\1AB} = O(r) . 
\end{split}
\end{equation}
There are two things to note here. First, the $O(r)$ coefficient of $h_{\1AB}$ corrects the corresponding free-field coefficient, so that
\begin{equation}
\begin{split}
\overset{-1,0}{h_{AB}}(u,{\hat x}) = \overset{-1,0}{h_{\0AB}}(u,{\hat x}) + \k \overset{-1,0}{h_{\1AB}}(u,{\hat x}) + O(\k^2) . 
\end{split}
\end{equation}
Indeed, this component receives corrections at all orders in $\k$. We can Fourier transform the trace-free part of the full coefficient as in \eqref{grav_rad_def} and the corresponding creation and annihilation operators are the $out$ operators. In summary,
\begin{equation}
\begin{split}
\label{grav_rad_def_out}
\overset{-1,0}{h_{\{AB\}}}(u,\hat{x}) &=  - \frac{i}{8\pi^2} \int_{0}^{\infty}\dt \o \left[  a^\text{out}_{AB} (\o ,\hat{x}) e^{-i\o u}-a_{AB}^{\text{out}\dagger} (\o ,\hat{x})e^{i\o u}\right] \equiv h_{AB}^\text{out}(u,{\hat x}) .
\end{split}
\end{equation}
Here and below, braces around angular indices denote the symmetric trace-free part with respect to $\g_{AB}$.

Secondly, we note that compared to the free field fall-offs \eqref{hfree}, several of the remaining metric components develop over-leading terms in the large-$r$ expansion. In particular, we have (see Appendix \ref{app:retarded_sol} for the derivation)
\begin{equation}
\begin{split}
\label{u-independent-components}
\overset{1,0}{h_{\1rr}}({\hat x}) &= - \frac{1}{8\pi} q^\a({\hat x}) P_{\0\a}  , \qquad \overset{0,0}{h_{\1rA}}({\hat x}) = \p_A \overset{1,0}{h_{\1rr}}({\hat x})  , \qquad \overset{-1,0}{h^A_{\1A}}({\hat x}) = ( D^2 + 1 )  \overset{1,0}{h_{\1rr}}({\hat x})  .
\end{split}
\end{equation}
$P_{\0\a}$ is the total 4-momentum passing through $\CI^+$ at $O(\k^0)$.

\paragraph{Solution at $O(\k^2)$:} At this order, the metric perturbation satisfies 
\begin{equation}
\label{second_order_eq_grav}
\n^2 \hb_{\2\mu\nu}=-\frac{1}{2} T_{\1\mu\nu} , \qquad T_{\1\mu\nu}  \equiv T^{\varphi}_{\1\mu\nu}+T^{\text{H}}_{\1\mu\nu} .  
\end{equation}
The retarded solution to this is similar to \eqref{h_ret_sol}
\begin{equation}
\begin{split}
\label{h_ret_sol_1}
{\bar h}_{\2\a\b}(X)  = \frac{1}{4\pi} \int \dt^4 X' \d((X-X')^2) \t(X^0-X'^0) T_{\1\a\b}(X') .
\end{split}
\end{equation}
As before, we use this to extract the leading large $r$ coefficients of $h_{\2\mu\nu}$. The key difference lies in the fact that the overleading fall-offs for the metric at $O(\k)$ induce overleading fall-offs in the stress tensor at $O(\k)$. In particular, substituting the fall-offs \eqref{hdd_falloffs_1} into \eqref{st_grav_leading}, we find
\begin{equation}
\label{grav_st_large_r_1}
\badat2
T_{\1uu}^h=O(r^{-2}),\qquad &T_{\1ur}^h=O(r^{-3}), \qquad T_{\1uA}^h=O(r^{-1}),
\\
T_{\1rr}^h=O(r^{-4}), \qquad & T_{\1rA}^h=O(r^{-2}),\qquad T_{\1AB}^h=O(r^0).
\eadat
\end{equation}
At this order, the scalar stress tensor retains the free-field fall-offs \eqref{scalar_st_large_r}, so the total stress tensor satisfies \eqref{grav_st_large_r_1}.
These fall-offs then induce overleading terms in certain metric components compared to \eqref{hdd_falloffs_1}. We show in Appendix \ref{app:retarded_sol} that
\begin{equation}
\begin{split}
\label{hdd_falloffs_2}
h_{\2uu} &= O(r^{-1} \ln (r/L) ) , \qquad h_{\2ur} = O(r^{-1}) , \qquad h_{\2uA} = O(\ln(r/L)) ,\\
h_{\2rr} &= O(r^{-1}) , \qquad \qquad \quad~ h_{\2rA} = O(r^0) , \qquad~ h_{\2AB} = O(r \ln(r/L)) , 
\end{split}
\end{equation}
with
\begin{equation}
\begin{split}
\label{second_order_results}
\overset{1,0}{h_{\2rr}}({\hat x}) &= - \frac{1}{8\pi} q^\a({\hat x}) P_{\1\a}  , \qquad\qquad \overset{0,0}{h_{\2rA}}({\hat x}) = \p_A \overset{1,0}{h_{\2rr}}({\hat x})  , \\
\overset{-1,0}{h^A_{\2A}}({\hat x}) &= ( D^2 + 1 )  \overset{1,0}{h_{\2rr}}({\hat x})   ,\qquad \overset{-1,1}{h_{\2AB}} (u,{\hat x}) = - \frac{1}{2} \overset{1,0}{h_{\1rr}}({\hat x}) \p_u \overset{-1,0}{h_{\0AB}} (u,{\hat x}) ,
\end{split} 
\end{equation}
where $P_{\1\a}$ is the energy-momentum through $\CI^+$ at $O(\k)$.

\paragraph{Generalisation to all orders in $\k$:} 

The key observation we make here is that the first three results in \eqref{second_order_results} preserve the structure of the $O(\k)$ results and are controlled by the total momentum through $\CI^+$. Additionally, the logarithmic correction to the radiative field (last result in \eqref{second_order_results}) is proportional to the $u$-derivative of the free radiative field. This derivative structure strongly suggests that the logarithmic term can be understood as a shift in $u$. In fact, it can be verified that such a structure continues also to $O(\k^3)$. Motivated by these results, we conjecture that to all orders in $\k$, the large-$r$ asymptotics for the metric coefficients are given by
\begin{equation}
\begin{split}
\label{all_order_res}
h_{rr}(u,r,{\hat x}) &= \frac{1}{r} \SP ({\hat x})  + o(r^{-1}) , \\
h_{rA}(u,r,{\hat x}) &=  D_A \SP ({\hat x})   + o(r^0)  , \\
\g^{AB} h_{AB}(u,r,{\hat x}) &= r  ( D^2 + 1 ) \SP ({\hat x})   + o(r)  , \\
h_{\{AB\}}(u,r,{\hat x}) &= r h^\text{out}_{AB} \left( u - \frac{\k}{2} \SP ({\hat x}) \ln \frac{r}{L}  , {\hat x} \right)  + o(r) ,
\end{split}
\end{equation}
where
\begin{equation}
\begin{split}
\SP ({\hat x}) \equiv - \frac{\k}{8\pi} q^\a({\hat x}) P_\a = \overset{1,0}{h_{rr}}({\hat x}).
\end{split}
\end{equation}
Here $P_\a$ is the exact energy-momentum passing through $\CI^+$. The leading $1/r$ coefficient of $h_{rr}$ will play a rather important role in the rest of this paper, so we have renamed $\overset{1,0}{h_{rr}} \to \SP$.

In the rest of this paper, we will assume \eqref{all_order_res}. In the follow-up work \cite{Choi:2026xxx}, we will prove it non-perturbatively in $\k$.

%%%%%%%%%%%%%%%%%%%%%%%%%%%%%%
\subsubsection{Scalar field}
%%%%%%%%%%%%%%%%%%%%%%%%%%%%%%

In the previous section, we used the retarded Green's function to argue that long range interactions modify the free-field fall-offs \eqref{hfree} with additional overleading terms at large $r$, e.g. \eqref{all_order_res}. Additionally, such interactions also modify the outgoing radiative gravitational field by inducing a shift in retarded time $u$ (see the last line of \eqref{all_order_res}). For the scalar field, long-range interactions do not introduce additional overleading power-law terms relative to the free $1/r$ fall-off \eqref{free_scalar_large_r}, but they do introduce overleading terms of the form $r^{-1} (\ln \frac{r}{L} )^n$. In this section, we show that the higher-order logarithmic terms can be resummed to all orders in \(\k\), producing the same shift of retarded time in the outgoing radiative scalar field. We emphasise that unlike the gravitational field, where \eqref{all_order_res} is conjectured for the purposes of this paper (proved in \cite{Choi:2026xxx}), we will derive the shift for the scalar field in this paper.

To show this, we shall shift tactics. Instead of using the retarded propagator and evaluating the large $r$ expansion order-by-order in $\k$, we will now work exactly in $\k$ and solve the scalar equations of motion asymptotically at large $r$. In other words, instead of treating $\k$ as a perturbative parameter, we will use $1/r$. The large $r$ asymptotics of the metric perturbation are given by \eqref{all_order_res}. For the scalar field, we make the following ansatz
\begin{equation}
\label{Scalar field ansatz}
\varphi(u,r,\hat{x})  = \frac{1}{r} \sum_{n=0}^{\infty} \left( \ln \frac{r}{L} \right)^n  \overset{1,n}{\varphi}(u,\hat{x}) + o(r^{-1}) . 
\end{equation}
We impose the boundary condition that the asymptotic scalar data $\varphi^\text{out}=\overset{1,0}{\varphi}$ vanishes as $\varphi^\text{out} = O(|u|^{-\e})$ for some $\e > 0$ as $u \to \pm\infty$. This is consistent with the assumptions made for the stress tensor in Appendix \ref{app:retarded_sol}.

Substituting this expansion into the scalar equations of motion \eqref{eom_scalar} and extracting the equations at each order in $r^{-2}(\ln \frac{r}{L} )^n$, we find the following recursive equations for the leading logarithmic coefficients,
\begin{equation}\label{recursion_rel}
\p_u\overset{1,n+1}{\varphi}(u,\hat x)=\frac{1}{(n+1)}\left( -\frac{\k}{2}\SP(\hat{x})\p_u\right) \p_u\overset{1,n}{\varphi}(u,\hat x) . 
\end{equation}
Importantly, this recursion relation is exact in $\k$. \eqref{recursion_rel} is solved by
\begin{equation}
\label{recursion_sol}
\overset{1,n}{\varphi}(u,\hat{x})=\frac{1}{n!}\left( -\frac{\k}{2}\SP(\hat{x}) \p_u \right) ^{n} \varphi^\text{out}(u,\hat{x}) . 
\end{equation}
The integration constants have been fixed by the boundary condition stated above.
As with the graviton field, we denote the exact log-free coefficient of $1/r$ by the superscript ``out''.\footnote{The log-free coefficient of $1/r$ at $\k=0$ denotes the incoming field, $\varphi^\text{in}$.} Using \eqref{recursion_sol}, we can resum the logarithms and we find that the asymptotic behaviour for the scalar field is
\begin{equation}\label{scalar_asymp_exact}
\varphi(u,r,\hat{x}) = \frac{1}{r} \varphi^\text{out} \left(u-\frac{\k}{2}\SP(\hat x) \ln \frac{r}{L}  ,\hat{x}\right)  + o(r^{-1}) ,
\end{equation}
which is the same logarithmic shift in the retarded time $u$ as in \eqref{all_order_res}.

Using this, we can evaluate the large $r$ expansion for the scalar stress tensor,
\begin{equation}
\label{Interacting stress tensor}
\badat{3}
T_{uu}^{\varphi}(u,r,\hat{x})&=\frac{1}{r^2} \overset{2,0}{T^\varphi_{uu}} \left(u-\frac{\k}{2}\SP(\hat{x})\ln \frac{r}{L} ,\hat{x}\right) +o(r^{-2}),
\\
T_{ur}^\varphi(u,r,\hat{x})&=\frac{1}{r^{3}} \left[-\frac{\k}{2}\SP(\hat{x})  \overset{2,0}{T^\varphi_{uu}} \left(u-\frac{\k}{2}  \SP(\hat{x})\ln \frac{r}{L} ,\hat{x}\right) \right]+o(r^{-3}),
\\
T_{uA}^{\varphi}(u,r,\hat{x})&=\frac{1}{r^2}\left[ \overset{2,0}{T^\varphi_{uA}} \left(u-\frac{\k}{2}\SP(\hat{x})\ln \frac{r}{L} ,\hat{x}\right)\right. \\
&\left. \qquad  \quad -\frac{\k}{2} \p_A \SP(\hat{x}) \overset{2,0}{T^\varphi_{uu}} \left(u-\frac{\k}{2}  \SP(\hat{x}) \ln \frac{r}{L} ,\hat{x}\right) \ln \frac{r}{L} \right] +o(r^{-2}),
\eadat
\end{equation}
where the stress-tensor components constructed from the outgoing radiative data are
\begin{equation}
\begin{split}
\overset{2,0}{T^\varphi_{uu}} = \left( \p_u \varphi^\text{out} \right)^2  , \qquad \overset{2,0}{T^\varphi_{uA}} = \p_u \varphi^\text{out} \p_A \varphi^\text{out} . 
\end{split}
\end{equation}

%%%%%%%%%%%%%%%%%%%%%%%%%%%%%%%%%%%%%%%%%%%%%%%%%%%%%%%%%%%%
\section{Large diffeomorphisms and Noether charges}
\label{sec:large_diffeos_charge}
%%%%%%%%%%%%%%%%%%%%%%%%%%%%%%%%%%%%%%%%%%%%%%%%%%%%%%%%%%%%

We now introduce the diffeomorphism vector fields compatible with the above asymptotics and the de Donder gauge conditions corresponding to supertranslations and superrotations. We then construct the associated Noether charges in the interacting theory and extract the regulated hard and soft charges.  We will first investigate the effect of non-local interactions on the Noether charge for supertranslations before studying superrotations which are our main focus.
%%%%%%%%%%%%%%%%%%%%%%%%%%%%%%%%%%%%%%%%%%%%%%%%%%%%%%%%%%%%
\subsection{Supertranslations and superrotations}
\label{sec:large_diffeos}
%%%%%%%%%%%%%%%%%%%%%%%%%%%%%%%%%%%%%%%%%%%%%%%%%%%%%%%%%%%%

Recall from \eqref{h_diff_transform} that the metric perturbation transforms under infinitesimal diffeomorphisms as
\begin{equation}
\begin{split}
\d_\xi h_{\mu\nu} = \frac{1}{\k} ( \n_\mu \xi_\nu + \n_\nu \xi_\mu ) + \CL_\xi h_{\mu\nu} . 
\end{split}
\end{equation}
In de Donder gauge \eqref{harmonic_gauge}, residual diffeomorphisms are infinitesimally generated by vector fields satisfying
\begin{equation}
\begin{split}
\label{residual_gauge_harmonic}
\n^2 \xi_\mu + \k [ h_{\mu\nu} \n^2 \xi^\nu + \n^{(\nu} \xi^{\rho)} ( 2 \n_\rho h_{\mu\nu} - \n_\mu h_{\nu\rho} ) ] = 0 . 
\end{split}
\end{equation}
We consider large diffeomorphisms that are non-vanishing on $\CI^+$ and behave as
\begin{equation}
\begin{split}
\xi = T(u,{\hat x}) \p_u + Y^A(u,{\hat x}) \p_A - r \a(u,{\hat x}) \p_r + \dots  ,
\end{split}
\end{equation}
where the $\dots$ denote subleading terms in $r$. Substituting this into \eqref{residual_gauge_harmonic} and simultaneously requiring that the diffeomorphism preserves the fall-offs, we find that residual large diffeomorphisms have the form\footnote{Recall that de Donder gauge does not completely fix all the residual small diffeomorphisms. Consequently, there are generically additional integration constants at subleading orders that we have suppressed here. These do not play a role in our analysis.}
\begin{equation}
\begin{split}
\xi = \left( T  + O(r^{-1}) \right) \p_u + \left( - r \a + \frac{1}{2} D^2 T + O \left( r^{-1} \ln\frac{r}{L} \right) \right) \p_r + \left( Y^A - \frac{1}{r} D^A T + O(r^{-2}) \right) \p_A ,
\end{split}
\end{equation}
where
\begin{equation}
\begin{split}
T(u,{\hat x}) = f({\hat x}) + u \a({\hat x}) , \qquad \a = \frac{1}{2} D \cdot Y,\qquad D_{\{A} Y_{B\}}({\hat x}) = 0  . 
\end{split}
\end{equation}
Here $f({\hat x})$ is a generic smooth function on $S^2$ that generates supertranslations. The associated vector field is
\begin{equation}
\begin{split}
\label{ST_vector_1}
\xi_f &= \left( f  + O(r^{-1}) \right) \p_u + \left(   \frac{1}{2} D^2 f + O \left( r^{-1} \ln\frac{r}{L} \right) \right) \p_r + \left(  - \frac{1}{r} D^A f + O(r^{-2}) \right) \p_A .
\end{split}
\end{equation}
$Y^A$ is a local conformal Killing vector on $S^2$ which, in stereographic coordinates, takes the form $Y^A \p_A = Y^z(z) \p_z + Y^\bz(\bz) \p_\bz$. The associated transformations are known as Virasoro superrotations generated by the vector field \begin{equation}
\begin{split}
\label{SR_vector_1}
\xi_Y &= \left( u \a   + O(r^{-1}) \right) \p_u + \left( - ( u + r ) \a  + O \left( r^{-1} \ln\frac{r}{L} \right) \right) \p_r + \left( Y^A - \frac{u}{r} D^A \a + O(r^{-2}) \right) \p_A  . 
\end{split}
\end{equation}

%%%%%%%%%%%%%%%%%%%%%%%%%%%%%%%%%%%%%%%%%%%%%%%%%%%%%%%%%%%%
\subsection{Noether charges}
\label{sec:Noether_charge}
%%%%%%%%%%%%%%%%%%%%%%%%%%%%%%%%%%%%%%%%%%%%%%%%%%%%%%%%%%%%

On a Cauchy slice $\S$, the charge associated to diffeomorphisms is given by
\begin{equation}
\begin{split}
\d Q_\xi[\S] = \int_\S \dt \S_\mu \o^\mu (\d , \d_\xi ) ,
\end{split}
\end{equation}
where $\o^\mu$ is the symplectic current and $\d_\xi$ is the variation of the fields under the diffeomorphism which acts on the fields $h$ and $\varphi$ as
\begin{equation}
\begin{split}
\d_\xi h_{\mu\nu} = \D_\xi h_{\mu\nu}  + \CL_\xi h_{\mu\nu} , \qquad \d_\xi \varphi = \CL_\xi \varphi , \qquad \D_\xi h_{\mu\nu} \equiv  \frac{1}{\k} ( \n_\mu \xi_\nu + \n_\nu \xi_\mu ) .
\end{split}
\end{equation}
We break up the transformation into two pieces $\D_\xi$ and $\CL_\xi$. In the asymptotic limit considered below, the charge associated with $\CL_\xi$ takes the form (see \cite{He:2020ifr})
\begin{equation}
\begin{split}
Q^\text{H}_\xi[\S] = - \int_\S \dt \S_\mu T^\mu{}_\nu \xi^\nu ,
\end{split}
\end{equation}
where $T_{\mu\nu}$ is the conserved stress tensor of the theory. We refer to this as the \emph{hard charge}. The charge associated to $\D_\xi$ is the \emph{soft charge}, which satisfies
\begin{equation}
\begin{split}
\d Q^\text{S}_\xi[\S] = \int_\S \dt \S_\mu \o^\mu (\d , \D_\xi ) . 
\end{split}
\end{equation}
The total charge is
\begin{equation}
\begin{split}
Q_\xi[\S] = Q_\xi^\text{S}[\Sigma] + Q_\xi^\text{H}[\Sigma] . 
\end{split}
\end{equation}

To construct the Noether charges for supertranslations and superrotations in the interacting theory, we start from the Lagrangian density
\begin{equation}
\begin{split}
\CL = \sqrt{-g} \left( \frac{2}{\k^2} \CR-\frac{1}{2}g^{\mu\nu}\p_\mu \varphi\p_\nu \varphi \right)  . 
\end{split}
\end{equation}
Varying this, we find
\begin{equation}
\d \CL = \CE_{\mu\nu} \d g^{\mu\nu} + \CE \d \varphi + \p_\mu \t^\mu(\d)   ,
\end{equation}
where $\CE_{\mu\nu}=0$ and $\CE=0$ are the graviton and scalar equations of motion \eqref{eom_grav} and \eqref{eom_scalar}, respectively. The pre-symplectic potential density is
\begin{equation}\label{symp_potential}
\t^\mu (\d) = \sqrt{-g}  \left( \frac{2}{\k^2} (g^{\nu\rho}\d \G^\mu _{\nu\rho}-g^{\mu\nu}\d \G^\rho_{\nu\rho} ) - g^{\mu\nu} \p_\nu \varphi \d \varphi \right) . 
\end{equation}
The symplectic current is defined to be
\begin{equation}
\label{symp_form_def}
\sqrt{-g} \o^\mu (\d,\d')=\d\t^\mu (\d')-\d'\t^\mu (\d)- \t^\mu ([\d,\d']).
\end{equation}
To evaluate the soft (hard) charges for supertranslations and superrotations we set, respectively, $\d' = \D_{\xi_f}$ ($\d' = \CL_{\xi_f}$) and $\d' = \D_{\xi_Y}$ ($\d' = \CL_{\xi_Y}$), integrate $\o$ over a Cauchy slice $\S_t$ and then take $t \to \infty$ (in units of $L$). Only the quadratic terms in the linearised gravity approximation contribute in this limit. We now study both cases in turn.

To deal with potential infrared divergences that may arise in our calculation, we introduce a long distance regulator $R$, evaluate the charge on a regulated Cauchy slice $\S_R$ at $t=R$ (or $r=R-u$) with $u \in ( - R , R )$, and then expand the charges around $R\gg L$.\footnote{We use the regulated Cauchy surface that has previously been used in \cite{Choi:2024ygx, Choi:2024mac, Choi:2024ajz}. However, it should be noted that in the large $R/L$ limit, the regulated slice $\S_R$ limits to $i^+ \cup \CI^+ \cup i^0_+$ where $i^0_+$ is the subset of $i^0$ with $t>0$. To land precisely on the asymptotic Cauchy slice $i^+ \cup \CI^+$, we need to take $u \in (-\rho(R),R)$ where $\rho(R)$ is a function that grows sublinearly as $R \to \infty$ (this was used in Appendix \ref{app:retarded_sol} to deduce the large $r$ expansion of the retarded solution). However, it seems that this choice for the regulated slice leads to the wrong soft superrotation charge. We hope to address this problem in future work.} As we will see shortly, these divergences are rather mild for the supertranslation charge considered in this subsection, but are more serious for superrotation charges studied in the next subsection.

%%%%%%%%%%%%%%%%%%%%%%%%%%%%%%%%%%%%%%%%%%%%%%%%%%%%%%%%%%%%
\subsubsection{Supertranslations}
\label{sec:ST_charge}
%%%%%%%%%%%%%%%%%%%%%%%%%%%%%%%%%%%%%%%%%%%%%%%%%%%%%%%%%%%%
In this subsection, we evaluate the Noether charge for supertranslations on the  asymptotic future Cauchy slice $\S^+ = \CI^+ \cup i^+$. This is evaluated via the limit,
\begin{equation}
\begin{split}
\label{Tf_def}
T^+_f \equiv \lim_{R \to \infty} Q_{\xi_f}[\S_R] .
\end{split}
\end{equation}
The large $R$ limit here is taken relative to the scale $L$.

%%%%%%%%%%%%%%%%%%%%%%%%%%%%%%%%%%%%%%%%%%
\subsubsection*{Soft charge}
%%%%%%%%%%%%%%%%%%%%%%%%%%%%%%%%%%%%%%%%%%

To evaluate the soft charge, we set $\d' = \D_{\xi_f}$ in \eqref{symp_form_def}, integrate $\o$ over the regulated Cauchy slice $\S_R$ and then take $R \to \infty$. The quadratic terms, which are the only ones that contribute in the linearised gravity approximation in this limit, are
\begin{equation}\label{omega_g^t}
\o^t(\d,\D_{\xi_f})=\frac{2}{\k} \d\left[\D_{\xi_f}\hb^{\a\b}\G_{\a\b}^{t}-\hb^{\a\b}\D_{\xi_f}\G^{t}_{\a\b}-\D_{\xi_f}\hb^{t\a}\G^\b_{\a \b}+\hb^{t\a}\D_{\xi_f}\G^\b_{\a\b}\right],
\end{equation}
where
\begin{equation}
\G^{t}_{\a\b}=\frac{\k}{2}(\eta^{u\s }+\eta^{r\s })\left(\n_\a h_{\s \b}+\n_\b h_{\s \a}-\n_\s h_{\a\b}\right),
\qquad \G^\b _{\a\b}=-\frac{\k}{2}\n_\a \hb.
\end{equation}
Using this and retaining the leading terms in the large $R$ limit, we then find
\begin{equation}
\label{pre_ST_gravity_charge}
\begin{split}
T^{+\rm{S}}_f(R) &=  -\frac{2}{\k}\int_{-R}^{R} \dt u \int_{S^2} \dt \O_2 D^AD^B f(\hat{x}) \p_u h^\text{out}_{\{AB\}}\left(u-\frac{\k}{2}\SP({\hat x})\ln \frac{R}{L},\hat{x}\right)  + o(R^0) \\
&= -\frac{2}{\k} \int_{S^2} \dt \O_2 D^AD^B f(\hat{x})  \int_{-R-\frac{\k}{2} \SP({\hat x})\ln\frac{R}{L}}^{R-\frac{\k}{2} \SP({\hat x})\ln\frac{R}{L}} \dt u \p_u h^\text{out}_{\{AB\}} ( u , {\hat x} )  + o(R^0)  . 
\end{split}
\end{equation}
In the second line, we have changed integration variable $u \to u + \frac{\k}{2}\SP({\hat x})\ln \frac{R}{L}$. We now consider the limit $R/L\to\infty$. To do this, we use the fact that at large $|u|$, the radiative field behaves as
\begin{equation}
\begin{split}
\label{CAB_largeu}
h^\text{out}_{\{AB\}}  (u,\hat{x}) \quad \xrightarrow{u/L \to \pm \infty} \quad \pm \frac{1}{2\k} \CJ^{(-1)}_{AB,+}(\hat{x})  + \frac{1}{u}  h^{\text{out},\1\pm}_{\{AB\}}({\hat x})  + O(|u|^{-1-\e}),
\end{split}
\end{equation}
for some $\e > 0$. Here, $\CJ^{(-1)}_{AB,+}(\hat{x})$ is the leading soft-graviton operator on $\CI^+$.\footnote{\label{footnote1}$h_{\{AB\}}^\text{out}$ in harmonic gauge is related to shear tensor $C_{AB}$ in Bondi gauge via $C_{AB} = h_{\{AB\}}^\text{out} - 2 D_{\{A} D_{B\}}C$ where $C$ is the supertranslation Goldstone mode. The transverse traceless part of the metric in harmonic gauge does \emph{not} include the Goldstone mode, which therefore does not appear in the large $|u|$ expansion in \eqref{CAB_largeu}.}

This large $|u|$ expansion was derived in \cite{Choi:2024ajz} using the retarded solution and stress tensor conservation.\footnote{It is expected that when using the Feynman propagator, the large $|u|$ expansion also contains terms proportional to $\ln\frac{|u|}{L}$ and $\frac{1}{u}\ln\frac{|u|}{L}$ \cite{Isen:2026xoc}.} For functions admitting such a large $|u|$ fall-off, it is easy to show that
\begin{equation}
\begin{split}
\label{limit-property}
\int_{-R-\frac{\k}{2} \SP({\hat x})\ln\frac{R}{L}}^{R-\frac{\k}{2} \SP({\hat x})\ln\frac{R}{L}} \dt u \p_u h^\text{out}_{\{AB\}} ( u , {\hat x} ) &= \int_{-\infty}^\infty \dt u \p_u h^\text{out}_{\{AB\}} ( u , {\hat x} ) + \CO(R^{-1}) = \frac{1}{\k} \CJ^{(-1)}_{AB,+}(\hat{x}) + \CO(R^{-1}) . 
\end{split}
\end{equation}
 It then follows that the soft charge is 
\begin{equation}
\begin{split}
T^{+\rm{S}}_f(R) = -\frac{2}{\k^2}\int_{S^2} \dt \O_2 D^A D^B f(\hat{x}) \CJ^{(-1)}_{AB,+}({\hat x}) + o(R^0).
\end{split}
\end{equation}

%%%%%%%%%%%%%%%%%%%%%%%%%%%%%%%%%%%%%%%%%%
\subsubsection*{Hard charge}
%%%%%%%%%%%%%%%%%%%%%%%%%%%%%%%%%%%%%%%%%%

Using \eqref{st_scalar_leading},  \eqref{st_grav_leading} and \eqref{ST_vector_1}, we find that the hard supertranslation charge is
\begin{equation}
\begin{split}
\label{pre_ST_matter_charge}
T_f^{+\text{H}}(R) &= - \int_{\S_R} \dt \S_\mu T^\mu{}_\nu \xi_f^\nu = \int_{-R}^R \dt u \int_{S^2} \dt \O_2 f(\hat{x}) \overset{2,0}{T_{uu}}\left(u-\frac{\k}{2}\SP(\hat{x})\ln \frac{R}{L},\hat{x}\right) + o(R^0) .
\end{split}
\end{equation}
As for the soft charge, we can now shift $u \to u + \frac{\k}{2} \SP({\hat x}) \ln \frac{R}{L}$ and take the large $R$ limit. The stress tensor falls off as $O(1/|u|^{2+\e})$ at large $|u|$ so \eqref{limit-property} applies and we find
\begin{equation}
\label{ST_matter_charge}
T_f^{+\text{H}}(R) = \int_{\CI^+}\dt u\dt \O_2 f(\hat{x}) \overset{2,0}{T_{uu}} \left(u,\hat{x}\right) + o(R^0). 
\end{equation}

%%%%%%%%%%%%%%%%%%%%%%%%%%%%%%%%%%%%%%%%%%
\subsubsection*{Total charge}
%%%%%%%%%%%%%%%%%%%%%%%%%%%%%%%%%%%%%%%%%%
We therefore find that the total supertranslation charge is 
\begin{equation}
    \begin{split}
        T_f^+(R) = \overset{0}{T}{}_f^+ + o(R^0)
    \end{split}
\end{equation}
where
\begin{equation}
\begin{split}
\overset{0}{T}{}_f^+ = -\frac{2}{\k^2}\int_{S^2} \dt \O_2 D^A D^B f(\hat{x}) \CJ^{(-1)}_{AB,+}({\hat x}) + \int_{\CI^+}\dt u\dt \O_2 f(\hat{x}) \overset{2,0}{T_{uu}} \left(u,\hat{x}\right) . 
\end{split}
\end{equation}
This charge agrees with the one derived in \cite{Strominger:2013jfa, He:2014laa}, which was done in the absence of long-range interactions. There it was shown that the total charge $\overset{0}{T}{}_f^\pm$ is conserved between $\CI^-$ and $\CI^+$ (i.e., $\overset{0}{T}{}_f^+ = \overset{0}{T}{}_f^-$) thus showing that supertranslations are a symmetry of gravitational scattering. The associated Ward identity is Weinberg's leading soft graviton theorem \cite{Weinberg:1965aa}. We have shown here explicitly that this result (both the supertranslation charge and, consequently, the leading soft-graviton theorem) is unaffected by long-range interactions.
%%%%%%%%%%%%%%%%%%%%%%%%%%%%%%%%%%%%%%%%%%%%%%%%%%%%%%%%%%%%
\subsubsection{Superrotations}
\label{sec:SR_charge}
%%%%%%%%%%%%%%%%%%%%%%%%%%%%%%%%%%%%%%%%%%%%%%%%%%%%%%%%%%%%
In this subsection, we evaluate the Noether charge for superrotations on the  asymptotic future Cauchy slice $\S^+ = \CI^+ \cup i^+$. This is evaluated via the limit,
\begin{equation}
\begin{split}
\label{JY_def}
J^+_Y \equiv \lim_{R \to \infty} Q_{\xi_Y}[\S_R] ,
\end{split}
\end{equation}
where $\S_R$ is a again the regulated Cauchy slice described previously. We will find that due to long-range interactions, the large $R/L$ limit is ill-defined for the superrotation charges. 

%%%%%%%%%%%%%%%%%%%%%%%%%%%%%%%%%%%%%%%%%%
\subsubsection*{Soft charge}
%%%%%%%%%%%%%%%%%%%%%%%%%%%%%%%%%%%%%%%%%%

To obtain the soft charge for superrotations $J_Y^{+\rm{S}}$, we have to simply replace $\D_{\xi_f} \to \D_{\xi_Y}$ in \eqref{omega_g^t}. Simplifying, we find
\begin{equation}
J_Y^{+\rm{S}}(R) = - \frac{1}{2\k} \int_{-R}^R \dt u \int_{S^2} \dt \O_2 S^{AB} ( u\p_u  - 1 ) h^\text{out}_{\{AB\}} \left(u - \frac{\k}{2} \SP(\hat{x}) \ln \frac{R}{L} , \hat{x}\right) + o(R^0) ,
\end{equation}
where
\begin{equation}
\begin{split}
S^{AB} \equiv 2D^{\{A} D^{B\}} D\cdot Y . 
\end{split}
\end{equation}
To proceed, we write
\begin{equation}
\begin{split}
( u\p_u  - 1 ) h^\text{out}_{\{AB\}}  = 2 u \p_u h^\text{out}_{\{AB\}} - \p_u \left( u h^\text{out}_{\{AB\}} \right) .
\end{split}
\end{equation}
Then,
\begin{equation}
\begin{split}
\label{delta_JY_1}
J_Y^{+\rm{S}}(R) &= -\frac{1}{\k} \int_{-R}^R \dt u \int_{S^2} \dt \O_2 S^{AB} u \p_u h^\text{out}_{\{AB\}} \left(u - \frac{\k}{2} \SP(\hat{x}) \ln \frac{R}{L} , \hat{x}\right) \\
&\qquad \qquad \qquad \qquad + \frac{1}{2\k} \int_{S^2} \dt \O_2 S^{AB} \left[ u h^\text{out}_{\{AB\}} \left(u - \frac{\k}{2} \SP(\hat{x}) \ln \frac{R}{L} , \hat{x}\right) \right]^{u=R}_{u=-R} + o(R^0) . 
\end{split}
\end{equation}
Using \eqref{CAB_largeu}, the second term in \eqref{delta_JY_1} simplifies to\footnote{In \cite{Choi:2024ajz}, the supertranslation Goldstone mode was included in $h_{\{AB\}}^\text{out}$ which led to linear in $R$ divergence in the soft superrotation charge, which was then removed by a counterterm. As discussed in Footnote \ref{footnote1} of the present paper, here the Goldstone does \emph{not} appear in $h_{\{AB\}}^\text{out}$, so there is no divergence to cancel and the limit in \eqref{no_lindiv_large_R} is finite.}
\begin{equation}
\begin{split}
\label{no_lindiv_large_R}
\left[ u h^\text{out}_{\{AB\}} \left(u - \frac{\k}{2} \SP(\hat{x}) \ln \frac{R}{L} , \hat{x}\right) \right]^{u=R}_{u=-R} &= h^{\text{out},\1+}_{\{AB\}}({\hat x}) - h^{\text{out},\1-}_{\{AB\}}({\hat x}) + o(R^0) \\
&= - \int_{-\infty}^\infty \dt u \p_u [ u^2 \p_u  h^\text{out}_{\{AB\}} ( u , {\hat x} ) ]  + o(R^0) . 
\end{split}
\end{equation}
For the first term in \eqref{delta_JY_1}, we proceed as before by changing integration variable $u \to u + \frac{\k}{2}\SP({\hat x})\ln \frac{R}{L}$, which gives
\begin{equation}
\begin{split}
J_Y^{+\rm{S}}(R) &= - \frac{1}{2} \ln \frac{R}{L} \int_{S^2} \dt \O_2 S^{AB} \SP(\hat{x})   \int_{-R-\frac{\k}{2}\SP({\hat x})\ln \frac{R}{L}}^{R-\frac{\k}{2}\SP({\hat x})\ln \frac{R}{L}} \dt u \p_u h^\text{out}_{\{AB\}} ( u , {\hat x} ) \\
&\qquad \qquad \qquad \qquad \qquad  - \frac{1}{\k} \int_{S^2} \dt \O_2 S^{AB}  \int_{-R-\frac{\k}{2}\SP({\hat x})\ln \frac{R}{L}}^{R-\frac{\k}{2}\SP({\hat x}) \ln \frac{R}{L}} \dt u u
\p_u h^\text{out}_{\{AB\}} ( u , {\hat x} )  \\
&\qquad \qquad \qquad \qquad \qquad - \frac{1}{2\k} \int_{\CI^+} \dt u \dt \O_2 S^{AB} \p_u [ u^2 \p_u  h^\text{out}_{\{AB\}} ( u , {\hat x} ) ]  + o(R^0) .
\end{split}
\end{equation}
The first term above is simplified using \eqref{limit-property}. The large $R$ limit of the second term is logarithmically divergent for functions that behave as \eqref{CAB_largeu} at large $|u|$. For the second term, the large $R$ limit is given by
\begin{equation}
\begin{split}
\label{limit-property-1}
\int_{-R-\frac{\k}{2}\SP({\hat x})\ln \frac{R}{L}}^{R-\frac{\k}{2}\SP({\hat x}) \ln \frac{R}{L}} \dt u u
\p_u h^\text{out}_{\{AB\}} ( u , {\hat x} ) &= \ln \frac{R}{L} \int_{-\infty}^\infty \dt u \p_u [ u^2 \p_u  h^\text{out}_{\{AB\}} ( u , {\hat x} ) ] \\
&\qquad \qquad - \int_{-\infty}^\infty \dt u \ln \frac{|u|}{L} \p_u [ u^2 \p_u  h^\text{out}_{\{AB\}} ( u , {\hat x} ) ] + o(R^0) . 
\end{split}
\end{equation}
Putting it all together, we find that at large $R/L$, the soft charge has a logarithmically divergent term and a finite term,
\begin{equation}
\begin{split}
\label{SR_soft_charge_1}
J_Y^{+\rm{S}}(R) &=  \ln \frac{R}{L} \int_{\CI^+} \dt u \dt \O_2 S^{AB} \left( - \frac{1}{\k} \p_u [ u^2 \p_u h^\text{out}_{\{AB\}} ( u , {\hat x} ) ]  - \frac{1}{2}  \SP(\hat{x})  \p_u h^\text{out}_{\{AB\}} ( u , {\hat x} ) \right)  \\
&\qquad \qquad \qquad  + \frac{1}{\k} \int_{\CI^+} \dt u \dt \O_2 S^{AB} \left( \ln \frac{|u|}{L} - \frac{1}{2} \right) \p_u [ u^2 \p_u  h^\text{out}_{\{AB\}} ( u , {\hat x} ) ]   + o(R^0). 
\end{split}
\end{equation}
It is convenient to define the subleading log soft-graviton operator and the subleading soft-graviton operators
\begin{equation}
\label{suprot_soft_operators}
\begin{split}
\CJ^{(\ln)}_{AB,+}(\hat{x}) &\equiv \k  \int_{-\infty}^{+\infty}\dt u \p_u (u^2\p_u h^\text{out}_{\{AB\}}  (u,\hat{x}) ) , \\
\CJ^{(0)}_{AB,+}(\hat{x}) &\equiv \k  \int_{-\infty}^{+\infty}\dt u \left( \ln\frac{|u|}{L} - \frac{1}{2} \right) \p_u [ u^2 \p_u  h^\text{out}_{\{AB\}} ( u , {\hat x} ) ] .
\end{split}
\end{equation}
Using this, we can write the soft superrotation charge as
\begin{equation}
\begin{split}
\label{SR_soft_charge}
J_Y^{+\rm{S}}(R) &= - \frac{1}{\k^2} \ln \frac{R}{L} \int_{S^2} \dt \O_2 S^{AB} \left[\CJ^{(\ln)}_{AB,+}(\hat{x}) + \frac{\k}{2}  \SP(\hat{x}) \CJ^{(-1)}_{AB,+}(\hat{x})  \right] + \frac{1}{\k^2} \int_{S^2}  \dt \O_2 S^{AB} \CJ^{(0)}_{AB,+}(\hat{x}) \\
&\qquad + o(R^0) . 
\end{split}
\end{equation}

%%%%%%%%%%%%%%%%%%%%%%%%%%%%%%%%%%%%%%%%%%
\subsubsection*{Hard charge}
%%%%%%%%%%%%%%%%%%%%%%%%%%%%%%%%%%%%%%%%%%

Using the same procedure as above, we find that the superrotation hard charge is
\begin{equation}
\begin{split}
\label{SR_hard_charge}
& J_Y^{+\text{H}}(R) = \frac{\k}{2} \ln \frac{R}{L} \int_{\CI^+} \dt u \dt \O_2 \left[ \frac{1}{2} D\cdot Y \SP(\hat x)  - Y^A D_A \SP (\hat x) \right] \overset{2,0}{T_{uu}} (u,{\hat x}) \\
 & + \int_{\CI^+} \dt u \dt \O_2  \left[ \frac{u }{2} D\cdot Y \overset{2,0}{T_{uu}} (u,{\hat x}) + Y^A \overset{2,0}{T_{uA}} (u,{\hat x}) - \frac{\k}{2} \left( \frac{1}{2} D\cdot Y \SP(\hat x)  - Y^A D_A \SP (\hat x) \right) \overset{2,0}{T_{uu}} (u,{\hat x}) \right] \\
 &  - \frac{2}{\k} \int_{S^2} \dt \O_2\left[  (Y^A+D^A D \cdot Y ) D^B \SP({\hat x})\CJ^{(-1)}_{AB,+}({\hat x}) - \frac{\k}{2} Y^A D_A \SP({\hat x}) \left( \overset{1,0}{h^{\text{fin},+}_{uu}}({\hat x}) - \overset{1,0}{h^{\text{fin},-}_{uu}}({\hat x}) \right) \right]  \\
 &+ o(R^0), 
\end{split}
\end{equation}
where
\begin{equation}
    \begin{split}
        \overset{1,0}{h^{\text{fin},\pm}_{uu}} = \lim_{u \to \pm \infty} \left( 1 - \ln \frac{|u|}{L} u \p_u \right) \overset{1,0}{h_{uu}}.
    \end{split}
\end{equation}
Here, $\overset{1,0}{h_{uu}}$ is logarithmically divergent at large $|u|$ (see \eqref{app:huu10}) and the projector above removes this divergence so the limit is finite. $\overset{1,0}{h^{\text{fin},\pm}_{uu}}$ is related via equations of motion to the energy flux through $\CI^\pm$ (see \eqref{app:huu10}). We will explore this relationship in more detail in the follow-up paper \cite{Choi:2026xxx}.

%%%%%%%%%%%%%%%%%%%%%%%%%%%%%%%%%%%%%%%%%%
\subsubsection*{Total charge}
%%%%%%%%%%%%%%%%%%%%%%%%%%%%%%%%%%%%%%%%%%
We thus find that the total superrotation charge is
\begin{equation}\label{SuperrotationChargeSplit}
J_Y^+(R) = \ln \frac{R}{L} \overset{\ln}{J_Y^+} + \overset{0}{J_Y^+}+ o(R^0) ,
\end{equation}
where
\begin{equation}
\begin{split}
\label{SR_charge_defs}
\overset{\ln}{J_Y^+} &\equiv - \frac{1}{\k^2} \int_{S^2} \dt \O_2 S^{AB} \! \left[\CJ^{(\ln)}_{AB,+}(\hat{x}) + \frac{\k}{2}  \SP(\hat{x}) \CJ^{(-1)}_{AB,+}(\hat{x})  \right] \\
&\qquad \qquad \qquad \qquad + \frac{\k}{2}  \int_{\CI^+} \dt u \dt \O_2 \left[ \frac{1}{2} D\cdot Y \SP(\hat x)  - Y^A D_A \SP (\hat x) \right] \overset{2,0}{T_{uu}} (u,{\hat x}) , \\
\overset{0}{J_Y^+} &\equiv \frac{1}{\k^2} \int_{S^2}  \dt \O_2 S^{AB} \CJ^{(0)}_{AB,+}(\hat{x}) \\
& - \frac{2}{\k} \int_{S^2} \dt \O_2 \left[ (Y^A+D^A D \cdot Y ) D^B \SP({\hat x}) \CJ^{(-1)}_{AB,+}({\hat x})   - \frac{\k}{2} Y^A D_A \SP({\hat x}) \left( \overset{1,0}{h^{\text{fin},+}_{uu}}({\hat x}) - \overset{1,0}{h^{\text{fin},-}_{uu}}({\hat x}) \right) \right]    \\
& + \int_{\CI^+} \!\!\! \dt u \dt \O_2 \! \left[ \frac{u }{2} D\cdot Y \overset{2,0}{T_{uu}} (u,{\hat x})   + Y^A \overset{2,0}{T_{uA}} (u,{\hat x}) - \frac{\k}{2} \left( \frac{1}{2} D\cdot Y \SP(\hat x)  - Y^A D_A \SP (\hat x) \right) \overset{2,0}{T_{uu}} (u,{\hat x})  \right] \! . 
\end{split}
\end{equation}

The total superrotation charge thus has a logarithmically divergent term with coefficient $\overset{\ln}{J_Y^+}$, and a finite term $\overset{0}{J_Y^+} $. The massive version of these charges was found in \cite{Choi:2024ygx, Choi:2024ajz}. In the next section, we will show that conservation of the logarithmic charge implies the logarithmic soft theorem, and we conjecture that the finite charge gives the classical part of the loop-corrected Cachazo-Strominger soft theorem \cite{Cachazo:2014fwa}.

%%%%%%%%%%%%%%%%%%%%%%%%%%%%%%
%%%%%%%%%%%%%%%%%%%%%%%%%%%%%%
\section{From conservation laws to soft theorems}
\label{sec:softthem_derivation}
%%%%%%%%%%%%%%%%%%%%%%%%%%%%%%

We are now in a position to establish the equivalence between the conservation of the superrotation charge in the presence of classical long-range interactions and the logarithmic (and loop-corrected subleading) soft graviton theorem.

\subsection{Superrotations and subleading soft graviton theorems}\label{Section on subleading soft theorems}

The regulated charge associated to a superrotation \eqref{SR_vector_1} can be expressed as
\begin{equation}\label{SuperrotationChargeSplit2}
J_Y^+ (R) = \ln \frac{R}{L} \overset{\ln}{J_Y^+} \left[\xi_{Y}\right]+\overset{0}{J_Y^+} \left[\xi_{Y}\right].
\end{equation}
In the following we will show that $\overset{\ln}{J_Y^+}$ is a symmetry of the S-matrix and that its conservation between $\CI^-$ and $\CI^+$ yields a Ward identity that is equivalent to the classical logarithmic soft graviton theorem coupled to massless matter. Moreover, we will see that $\overset{0}{J_Y^+}$ truncated at leading order in the gravitational coupling is associated to the (tree-level) Cachazo-Strominger soft theorem, and interpret the corrections at higher order in $\k$ as 1-loop exact corrections to this soft theorem.
\\
For simplicity, in the following sections, we restrict ourselves to the case where all the hard external states are scalar modes.

%%%%%%%%%%%%%%%%%%%%%%%%%%%%%%%%%%%%%%%%%
\subsubsection{Logarithmic soft theorem}
%%%%%%%%%%%%%%%%%%%%%%%%%%%%%%%%%%%%%%%%%
The soft and hard contributions to the `logarithmic superrotation charge', $\overset{\ln}{J_Y^+}=\overset{\ln}{J_Y^{+\text{S}}}+\overset{\ln}{J_Y^{+\text{H}}}$, on the future null boundary $\CI^+$ read off from \eqref{SR_charge_defs} are given by
\begin{equation}
\label{Log charges}
\badat2
\overset{\ln}{J_Y^{+\text{S}}}&=\int\dt \O_2\  S^{AB}\left[-\frac{1}{\k^2}\CJ_{AB,+}^{(\ln)}-\frac{1}{2\k}\SP(\hat{x})\CJ_{AB,+}^{(-1)}\right],
\\
\overset{\ln}{J_Y^{+\text{H}}}&=\int\dt u\dt \O_2\ \left[\frac{\k}4(D\cdot Y)\SP(\hat{x})\overset{2,0}{T^{\varphi}_{uu}}(u,\hat{x})-\frac{\k}{2}Y^{A}(\p_A\SP(\hat{x}))\overset{2,0}{T^{\varphi}_{uu}}(u,\hat{x})\right],
\eadat
\end{equation}
where $\CJ_{AB,+}^{(-1)}$ is the leading soft graviton operator defined in \eqref{limit-property} and the `log soft graviton operator' $\CJ_{AB,+}^{(\ln)}$   \cite{Choi:2024ajz} has been defined in \eqref{suprot_soft_operators}.
Analogous charges $\overset{\ln}{J_Y^{-\text{S}}}$ and $\overset{\ln}{J_Y^{-\text{H}}}$ are obtained on the past null boundary $\CI^-$.
Upon antipodal identification of the radiation and matter fields as well as the symmetry parameter $Y$ across $i^0$ these charges are conserved
\begin{equation}
\overset{\ln}{J_Y^+}=\overset{\ln}{J_Y^-}.
\end{equation}
In the following we will show that this conservation law can be recast as a symmetry of the $\Scal$-matrix,
\begin{equation}
\langle {\rm out}|\overset{\ln}{J_Y^+} \Scal-\Scal \overset{\ln}{J_Y^-}|{\rm in}\rangle=0,
\end{equation}
and is equivalent to the logarithmic soft graviton theorem coupled to massless matter.

To connect the superrotation charge which is smeared on the celestial sphere to the logarithmic soft theorem,
we make a judicious choice of vector field $Y=(Y^z,Y^\zb)$ on the celestial sphere
\begin{equation}\label{Choice of the superrotation field}
Y^{z}=\frac{1}{z-w},\quad Y^{\zb}=0,
\end{equation}
which satisfies
\begin{equation}S_{zz}=2D_{z}^{3}Y^{z}=-\frac{12}{(z-w)^4}\ , \quad S_{\zb\zb}=0.
\end{equation}
With this, the soft logarithmic charge \eqref{Log charges} becomes
\begin{equation}
\badat{3}
\overset{\ln}{J_Y^{+\text{S}}}&=\frac{1}{\k^2}\int\dt ^2z \g_{z\zb} \ \frac{12(\g^{z\zb})^2}{(z-w)^4}\CJ^{(\ln)}_{\zb\zb}+\frac{1}{\k}\int\dt ^2z \g_{z\zb}\ \frac{6(\g^{z\zb})^2}{(z-w)^4} \SP(\hat{x}_{z})\CJ^{(-1)}_{\zb\zb},
\\
&=\frac{3i}{2\pi\k}\int\dt ^2z\ \frac{1}{(z-w)^4}\lim_{\o\to 0+}\p_{\o}\o^2\p_{\o}\left[ a_{-}(\o\hat{x}_{z})-a^{\dagger}_{+}(\o\hat{x}_{z}) \right]
\\
&-\frac{3}{4\pi}\int\dt ^2z\ \frac{\SP(\hat{x}_{z})}{(z-w)^4}\lim_{\o\to 0+}\o\left[a_{-}(\o\hat{x}_{z})+a^{\dagger}_{+}(\o\hat{x}_{z})\right].\eadat
\end{equation}
The Coulombic term involving $\SP$ can be obtained by solving the Einstein equations with retarded boundary conditions, and is given by 
\begin{equation}
\label{Writing h_rr(1,0) as an integral over the stress tensor}
\SP(\hat{x})=-\frac{\k}{8\pi}\int_{S^2}\dt \O_2'\int_{-\infty}^{\infty}\dt u'\left(q\cdot q'\right)\overset{2,0}{T^{\varphi}_{u'u'}}(u',\hat{x}').
\end{equation}
Given an arbitrary $\bra{\text{out}}$ Fock state and a function $f(\hat{x})$, we have that
\begin{equation}
\label{Action of the integrated stress tensor on out states}
\bra{\text{out}}:\int_{S^2}\dt \O_2\int_{-\infty}^{\infty}\dt u\ \overset{2,0}{T^{\varphi}_{uu}}(u,\hat{x}) f(\hat{x})\ :=\sum_{i\in\text{out}}\o_{i} f(\hat{x}_{i})\bra{\text{out}}.
\end{equation}
A similar analysis can be carried out on the past boundary. With this the contribution from the soft logarithmic charge to the Ward Identity can be written as 
\begin{equation}
\label{Contribution of the soft charges to the log ward identity}
\badat2
\bra{\text{out}}\overset{\ln}{J_Y^{+\text{S}}}\Scal-\Scal\overset{\ln}{J_Y^{-\text{S}}}\ket{\text{in}}=&\int\dt ^2z \ \frac{3i}{2\pi\k}\frac{1}{(z-w)^4}\lim_{\o\to 0}\p_{\o}\o^2\p_{\o}\bra{\text{out}}a_{-}(\o\hat{x}_{z})\Scal-\Scal a_{+}^{\dagger}(\o\hat{x}_{z})\ket{\text{in}} 
\\
&+\int\dt ^2z \  \frac{3\k^2}{32\pi^2}\frac{1}{(z-w)^4}\sum_{i\in\text{out}}\sum_{j\in\text{in,out}} (k\cdot p_{i})\frac{(\e^{-}\cdot p_{j})^2}{p_{j}\cdot k}\bra{\text{out}}\Scal\ket{\text{in}},
\eadat
\end{equation}
where $k=\o q(w,\wb)$ is the momentum of the soft graviton and we used momentum conservation $\sum_{i\in {\rm in},{\rm out}}p_i=0$ to replace the sum over incoming momenta in terms of outgoing ones. 
\\
Next we determine the action of the hard logarithmic charge on the asymptotic states. For the out state we have
\begin{equation}
\badat{3}
\bra{\text{out}}\overset{\ln}{J_Y^{+\text{H}}}&=\bra{\text{out}}: \int \dt u\dt \O_2\left[\frac{\k}4\SP(\hat x)( D\cdot Y)\overset{2,0}{T^{\varphi}_{uu}}(u,\hat{x})-\frac{\k}{2}Y^{A}\p_A\SP\overset{2,0}{T^{\varphi}_{uu}}(u,\hat{x})\right],
\\
&=\bra{\text{out}}:\int\dt u \dt \O_2 \dt u'\dt \O_2'\ \frac{\k^2}{16\pi}\overset{2,0}{T^{\varphi}_{uu}}(u,\hat{x})\left[-\frac{1}{2}(D\cdot Y)+Y^{A}\p_A\right]\left(q\cdot q'\right)\overset{2,0}{T^{\varphi}_{u'u'}}(u',\hat{x}'),
\\
&=\sum_{i,j\in\text{in,out}}\ \frac{\k^2}{16\pi}\left[-\frac{1}{2}(D\cdot Y)+Y^{A}\p_A\right]_{\hat{x}_{z_{i}}}\left(p_{i}\cdot p_{j}\right)\bra{\text{out}}.
\eadat
\end{equation}
A similar analysis can be performed for the action of the hard charge on incoming states. With this, and with the specific choice \eqref{Choice of the superrotation field} of our symmetry vector field, the total contribution of the hard log charges to the Ward identity is
\begin{equation}
\label{Contribution of the hard charges to the log ward identity}
\badat2
\bra{\text{out}}\overset{\ln}{J_Y^{+\text{H}}}\Scal-\Scal\overset{\ln}{J_Y^{-\text{H}}}\ket{\text{in}}=\sum_{i,j\atop\eta_{i}\eta_{j}=1}\frac{\k^2}{16\pi}\frac{3z_{i}-z_{j}-2w}{2(z_{i}-w)^2(z_{i}-z_{j})}\left(p_{i}\cdot p_{j}\right) \bra{\text{out}}\Scal\ket{\text{in}}.
\eadat
\end{equation}
The full Ward identity thus is
\begin{equation}\label{Full Log Ward Identity}
\badat{3}
0=&\int\dt ^2z \ \frac{3i}{2\pi\k}\frac{1}{(z-w)^4}\lim_{\o\to 0}\p_{\o}\o^2\p_{\o}\bra{\text{out}}a_{-}(\o\hat{x}_{z})\Scal -\Scal a_{+}^{\dagger}(\o\hat{x}_{z})\ket{\text{in}} 
\\
&+\int\dt ^2z \  \frac{3\k^2}{32\pi^2}\frac{1}{(z-w)^4}\sum_{i\in\text{out}}\sum_{j\in\text{in,out}} (k\cdot p_{i})\frac{(\e^{-}\cdot p_{j})^2}{p_{j}\cdot k}\bra{\text{out}}\Scal\ket{\text{in}},
\\
&+\sum_{i,j\atop\eta_{i}\eta_{j}=1}\frac{\k^2}{16\pi}\frac{3z_{i}-z_{j}-2w}{2(z_{i}-w)^2(z_{i}-z_{j})}\left(p_{i}\cdot p_{j}\right) \bra{\text{out}}\Scal\ket{\text{in}}.
\eadat
\end{equation}
It is known that using a local CKV superrotation (i.e. one that is parameterised by a meromorphic vector field of the form that we have chosen), the Ward identity does not map onto the soft theorem, but rather its shadow transform. 
\\
Performing a shadow transform\footnote{The shadow transform kernel here is proportional to $\frac{(x-w)^2}{(\xb-\wb)^2}$.} on the expression \eqref{Full Log Ward Identity}, we get
\begin{equation}
\badat2
0&=\frac{\k^2}{8} \sum_{i,j\atop\eta_{i}\eta_{j}=1}\frac{(x-z_{i})(x-z_{j}) }{(\xb-\zb_{i})(z_{i}-z_{j})}  \ (p_{i}\cdot p_{j})\bra{\text{out}}\Scal\ket{\text{in}} + \frac{\k^2}{8}\sum_{i\in\text{out}}\sum_{j\in\text{in,out}} (k\cdot p_{i})\frac{(\e^{-}\cdot p_{j})^2}{p_{j}\cdot k}\bra{\text{out}}\Scal\ket{\text{in}}\nonumber
\\
&+\frac{2i\pi}{\k}\lim_{\o\to 0}\p_{\o}\o^2\p_{\o}\bra{\text{out}}a_{-}(\o\hat{x}_{x})\Scal-\Scal a_{+}^{\dagger}(\o\hat{x}_{x})\ket{\text{in}}.
\eadat
\end{equation}
The normalisation of the shadow transform drops out since the whole expression is equated to zero.
\\
Rearranging terms gives us
\begin{equation}
\label{Log soft theorem, almost}
\badat2
\lim_{\o\to 0}\p_{\o}\o^2\p_{\o}\bra{\text{out}}a_{-}(\o\hat{x}_{x})\Scal -\Scal a_{+}^{\dagger}(\o\hat{x}_{x})\ket{\text{in}}&=\frac{i\k^{3}}{16\pi} \sum_{i,j\atop\eta_{i}\eta_{j}=1} \frac{(x-z_{i})(x-z_{j}) }{(\xb-\zb_{i})(z_{i}-z_{j})} (p_{i}\cdot p_{j}) \bra{\text{out}}\Scal\ket{\text{in}}
\\
&+ \frac{i\k^{3}}{16\pi}\  \sum_{i\in\text{out}}\sum_{j\in\text{in,out}} (k\cdot p_{i})\frac{(\e^{-}\cdot p_{j})^2}{p_{j}\cdot k}\bra{\text{out}}\Scal\ket{\text{in}}.
\eadat
\end{equation}
The first term in the R.H.S. of \eqref{Log soft theorem, almost} can be recognised as the following combination of external kinematical variables
\begin{equation}\label{}
\frac{(x-z_{i})(x-z_{j})}{(\xb-\zb_{i})(z_{i}-z_{j})}\left(p_{i}\cdot p_{j}\right)=-\frac{\e^{-}_{\mu\a}p_{i}^\a k_\nu }{p_{i}\cdot k}\left\{p_{i}^\mu \frac{\p}{\p p_{i\,\nu}}-p_{i}^\nu \frac{\p}{\p p_{i\,\mu}}\right\}\left(p_{i}\cdot p_{j}\right),
\end{equation}
where $k^\mu =\o q^\mu (\hat{x}_{x})$. With this, we can write the shadow-transformed Ward identity of meromorphic superrotations as 
\begin{equation}
\label{Log soft theorem,  using Virasoro + Shadow}
\badat2
&\lim_{\o\to 0}\p_{\o}\o^2\p_{\o}\bra{\text{out}}a_{-}(\o\hat{x}_{x})\Scal-\Scal a_{+}^{\dagger}(\o\hat{x}_{x})\ket{\text{in}}
\\
=&-\frac{i\k^{3}}{16\pi} \sum_{i,j\atop\eta_{i}\eta_{j}=1} \frac{\e^{-}_{\mu\a}p_{i}^\a k_\nu }{p_{i}\cdot k}\left\{p_{i}^\mu \frac{\p}{\p p_{i\,\nu}}-p_{i}^\nu \frac{\p}{\p p_{i\,\mu}}\right\}\left(p_{i}\cdot p_{j}\right)\bra{\text{out}}\Scal\ket{\text{in}}
\\
&+ \frac{i\k^{3}}{16\pi} \sum_{i\in\text{out}}\sum_{j\in\text{in,out}} (k\cdot p_{i})\frac{(\e^{-}\cdot p_{j})^2}{p_{j}\cdot k}\bra{\text{out}}\Scal\ket{\text{in}}.
\eadat
\end{equation}
This is exactly the logarithmic soft graviton theorem in a theory with massless matter coupled to gravity which can be derived from the massless limit of the result in \cite{Sahoo:2018lxl}. We have thus shown that superrotations are symmetries of the $\Scal$-matrix when long-range interactions are included, and that the conservation of a novel logarithmic superrotation charge is equivalent to the logarithmic soft graviton theorem in Einstein gravity coupled to massless matter.

%%%%%%%%%%%%%%%%%%%%%%%%%%%%%%%%%%%%%%%%%
\subsubsection{Subleading powerlaw soft theorem}%{Cachazo-Strominger soft theorem with one-loop exact corrections}
%%%%%%%%%%%%%%%%%%%%%%%%%%%%%%%%%%%%%%%%%
The soft and hard contributions to the superrotation charge $\overset{0}{J_Y}=\overset{0}{J_Y^{\text{S}}}+\overset{0}{J_Y^{\text{H}}}$ read off from \eqref{SR_charge_defs}   are given by
\begin{equation}
\label{Corrected Cachazo-Strominger charges}
\badat3
&\overset{0}{J_Y^{+S}} = \frac{1}{\k^2}\int_{S^2}\dt \O_2\  S^{AB}\CJ^{(0)}_{AB,+},
\\
&\overset{0}{J_Y^{+H}} =  - \frac{2}{\k} \int_{S^2} \!\!\!\! \dt \O_2 \! \left[ \! (Y^A+D^A D \cdot Y ) D^B \SP({\hat x}) \CJ^{(-1)}_{AB,+}({\hat x})   - \frac{\k}{2} Y^A D_A \SP({\hat x}) \left( \overset{1,0}{h^{\text{fin},+}_{uu}}({\hat x}) - \overset{1,0}{h^{\text{fin},-}_{uu}}({\hat x}) \right) \right]    \\
&\quad + \int_{\CI^+} \!\!\! \dt u \dt \O_2 \! \left[ \frac{u }{2} D\cdot Y \overset{2,0}{T_{uu}} (u,{\hat x})   + Y^A \overset{2,0}{T_{uA}} (u,{\hat x}) - \frac{\k}{2} \left( \frac{1}{2} D\cdot Y \SP(\hat x)  - Y^A D_A \SP (\hat x) \right) \overset{2,0}{T_{uu}} (u,{\hat x})  \right],
\eadat
\end{equation}
where $\CJ^{(0)}_{AB,+}(\hat{x})$ is the subleading soft graviton operator as defined in \eqref{suprot_soft_operators} while $\CJ_{AB,+}^{(-1)}$ is the leading soft graviton operator as defined in \eqref{limit-property}. Analogous charges $\overset{0}{J_Y^{-\text{S}}}$ and $\overset{0}{J_Y^{-\text{H}}}$ are obtained on the past null boundary.
In the absence of the Coulombic terms in \eqref{Corrected Cachazo-Strominger charges} involving $\SP\equiv\overset{1,0}{h_{rr}}$, these charges have previously appeared in \cite{Kapec:2014opa} where it was shown that the associated Ward identity yields the  (shadow-transform of) the subleading tree-level soft graviton theorem identified by Cachazo and Strominger in \cite{Cachazo:2014fwa}. Long-range interactions correct this charge as in \eqref{Corrected Cachazo-Strominger charges} and we interpret this $\k$-exact term as a 1-loop exact correction to the Cachazo-Strominger soft theorem \cite{Cachazo:2014fwa}. In our follow-up work \cite{Choi:2026xxx}, we will further discuss these corrections and interpret them in the context of celestial holography.

%%%%%%%%%%%%%%%%%%%%%%%%%%%%%%%%%%%%%%%%%%%%%%%%%%%%%%%%%%%%
\section{Tail memory}
\label{sec:memory}
%%%%%%%%%%%%%%%%%%%%%%%%%%%%%%%%%%%%%%%%%%%%%%%%%%%%%%%%%%%%
The core idea behind an infrared triangle  is a three-way equivalence between asymptotic symmetries, soft theorems and memory effects. The earlier sections have detailed the equivalence between asymptotic symmetries and soft theorems in presence of long-range gravitational interactions. In this section, we look at the corresponding memory effects.

\subsection*{Displacement memory}

At leading order in $\k$, the Coulombic (denoted with a superscript $C$) part of the transverse traceless metric on $\CI^+_+$ and $\CI^+_-$ is given by (see equation \eqref{hab_TT})
\begin{equation}\label{C_AB(0) using retarded propagator}
h_{\1\{AB\}}^{C,(0)+}{}(\hat{x})=-\frac{1}{8\pi}\int\dt \O_2'\dt u'\ \frac{(D_{\{A}q \cdot q')(D_{B\}}q \cdot q')}{q\cdot q'}\ \overset{2,0}{T_{\0u'u'}}(u',\hat{x}'),\qquad h^{C,(0)-}_{\1\{AB\}}{}=0,
\end{equation}
We have seen in Section \ref{sec:asymp_data} that higher-order corrections in $\k$ simply shift the $u'$ coordinate of the leading stress tensor $ u' \to u' - \frac{\k}{2} \SP({\hat x})\ln\frac{R}{L}$. We can remove this shift in the integral above by changing integration variables $u\ \to u\ + \frac{\k}{2} \SP({\hat x})\ln\frac{R}{L}$. The exact expression for the metric perturbation near $\CI^+_\pm$ is then given by
\begin{equation}
h_{\{AB\}}^{C,(0)+}{}(\hat{x})=-\frac{\k}{8\pi}\int\dt \O_2'\dt u'\ \frac{(D_{\{A}q \cdot q')(D_{B\}}q \cdot q')}{q\cdot q'}\ \overset{2,0}{T_{u'u'}}(u',\hat{x}'),\qquad h^{C,(0)-}_{\{AB\}}{}=0.
\end{equation}
A similar expression can be obtained on $\CI^-$. The precise forms of $h^C_{\{AB\}}$ on $\CI^+_+$ and $\CI^+_-$ is gauge-dependent. However, the difference is not! Indeed, the gauge-invariant combination
\begin{equation}\label{Displacement Memory}
\Delta(\k h_{\{AB\}}^{C,(0)}) =\kappa\left(h_{\{AB\}}^{C,(0)+}   - h_{\{AB\}}^{C,(0)-} \right)   
\end{equation}
is precisely the {\it linear displacement memory} \cite{Strominger:2014pwa}. Typically the linear displacement memory effect is given in terms of the asymptotic momenta of point particles. For massless point particles, $\mathcal{P}^\mu \sim q^\mu \overset{2,0}{T_{uu}}$ with the stress tensor localised onto a specific spacetime null-geodesic, so we have 
\begin{equation}
\Delta(\k h_{\{AB\}}^{C}) \sim -\frac{\k^2}{8\pi}\int\dt u'\dt \O_2' \frac{\left(\p_{\{A}q (\hat{x}) \cdot \mathcal{P}\right)\left(\p_{B\}}q (\hat{x})\cdot \mathcal{P}\right)}{\left(q(\hat{x})\cdot \mathcal{P}\right)}.
\end{equation}
This formula is the massless version of (3.39) in \cite{Choi:2024ajz} where the linear displacement memory formula in terms of asymptotic
momenta of massive particles \cite{Braginsky:1987kwo} was generalised to massive scalar fields. \eqref{Displacement Memory} is the analogous generalisation from massless particles to fields.

\subsection*{Tail to the displacement memory}
Long-range interactions create a {\it tail} to the displacement memory which at late times $u\to \infty$ is encoded in the $1/u$ term in \eqref{CAB_largeu}. Using similar arguments as above yields 
\begin{equation}
\label{C_AB(1) using retarded propagator}
\badat3
 h^{C,(1)+}_{\{AB\}}({\hat x}) &=\frac{\k}{8\pi}\int\dt \O_2'\dt u'\ \Big[D_{\{A}q^\mu D_{B\}}q^\nu D^{C}(q_\mu 'q_\nu ')\Big(\frac{\k}{2}D_{C}\SP(\hat{x}')\overset{2,0}{T_{u'u'}}(u',\hat{x}')+\overset{2,0}{T_{u'C'}}(u',\hat{x}')\Big)
\\
&\qquad+D_{\{A}q^\mu D_{B\}}q^\nu (D^{C}q'_\mu D^{D}q'_\nu )\overset{1,0}{T_{CD}}(u',\hat{x}')+(D_{\{A}q\cdot q')(D_{B\}}q\cdot q')\overset{3,0}{T_{u'u'}}(u',\hat{x}')\Big],
\\
h^{C,(1)-}_{\{AB\}}(\hat{x})&=0.
\eadat
\end{equation}
Similar expressions hold on $\CI^-$. We propose that
\begin{equation}
\Delta  \left(\kappa h^{C,\1}_{\{AB\}}({\hat x})\right) = \kappa\left(h^{C,\1+}_{\{AB\}}({\hat x}) - h^{C,\1-}_{\{AB\}}({\hat x})\right) 
\end{equation}
measures the `tail' to the displacement memory for a massless scalar field.

%%%%%%%%%%%%%%%%%%%%%%%%%%%%%%%%%%%%%%%%%%%%%%%%%%%%%%%%%%%%
\section*{Acknowledgements}
%%%%%%%%%%%%%%%%%%%%%%%%%%%%%%%%%%%%%%%%%%%%%%%%%%%%%%%%%%%%
We thank Federico Capone, Richard M. Myers, and Biswajit Sahoo for useful discussions. SC, PM and AP were supported by the European Research Council (ERC) under the European Union’s Horizon 2020 research and innovation programme (grant agreement No 852386). SC and PM were also supported by the European Research Council under the European Union’s Seventh Framework Programme (FP7/2007-2013),
ERC Grant agreement ADG 834878. PM is also supported by the STFC Consolidated
Grant ST/C004240/1.
SC is also supported by the European Research Council (ERC) under the European Union’s Horizon 2020 research and innovation programme (grant agreement No 852386), and the ERC Consolidator/UKRI Frontier grant TwistorQFT EP/Z000157/1.
This work was supported by the Simons Collaboration on Celestial Holography. The authors acknowledge use of ChatGPT and Claude for verification of several formulas throughout this paper. No scientific ideas, calculations, results, or conclusions were generated by ChatGPT or Claude.

%%%%%%%%%%%%%%%%%%%%%%%%%%%%%%%%%%%%%%%%%%%%%%%%%%%%%%%%%%%%
\appendix
%%%%%%%%%%%%%%%%%%%%%%%%%%%%%%%%%%%%%%%%%%%%%%%%%%%%%%%%%%%%

\section{Asymptotic expansion of the retarded solution}
\label{app:retarded_sol}

In this Appendix, we determine the large $r$ expansion of \eqref{h_ret_sol}, which we recall here
\begin{equation}
\begin{split}
\label{app:h_ret_sol_appendix}
{\bar h}_{\1\a\b}(X)  = \frac{1}{4\pi} \int \dt^4 X' \d((X-X')^2) \t(X^0-X'^0) T_{\0\a\b}(X') .
\end{split}
\end{equation}
We now express the Cartesian components as functions of retarded coordinates using \eqref{qt_def} and find
\begin{equation}
\begin{split}
{\bar h}_{\1\a\b}(u,r,{\hat x}) &= \frac{1}{4\pi}  \int_{-\infty}^\infty  \dt u' \int_0^\infty  \dt r' r'^2 \int_{S^2} \dt \O_2'\t(u-u'+r-r') T_{\0\a\b}(u',r',{\hat x}') \\
&\qquad \qquad \qquad \qquad \times \d((u-u')(u-u'+2r-2r')  + 2 r r' q({\hat x}) \cdot q({\hat x}') )  .
\end{split}
\end{equation}
We leave the Cartesian indices $\a\b$ as is, but change the arguments of all the functions to retarded coordinates. Next, we change integration variables $u' \to s=u-u'$ and $x'^A \to (\l,\phi)$, where $\l \equiv-q({\hat x})\cdot q({\hat x}')\in[0,2]$ and $\phi$ is the azimuthal angle about the observation direction specified by $q^\a({\hat x})$. The Dirac delta function then fixes $\l = \frac{s(s+2r-2r')}{2 r r'}$. The integrand contributes only if $\frac{s(s+2r-2r')}{2 r r'} \in [0,2]$ and $s+r-r' \geq 0$. These inequalities are equivalent to $s \geq 0$ and $\frac{s}2 \leq r' \leq \frac{s}2+r$. We then have
\begin{equation}
\begin{split}
{\bar h}_{\1\a\b} (u,r,{\hat x}) &= \frac{1}{8\pi r}  \int_0^\infty  \dt s \int_{\frac{s}2}^{\frac{s}2+r}  \dt r' r' \int_0^{2\pi}  \dt \phi T_{\0\a\b} \left( u - s , r' , \frac{s(s+2r-2r')}{2 r r'} , \phi \right) .
\end{split}
\end{equation}
We now consider the large $r$ limit of this quantity. Apart from the explicit factor of $1/r$, $r$ appears both in the upper limit of the $r'$-integration and in the angular argument of the stress tensor. The limit is potentially non-uniform because the integration domain always contains a region in which $r'$ and $s$ are comparable to $r$. To deal with these issues carefully, we break up the integral in ${\bar h}_{\1\a\b} = {\bar h}_{\1\a\b}^< + {\bar h}_{\1\a\b}^>$  into two regions $r'< \rho$ and $r'> \rho$ with $0<\rho<r$, yielding
\begin{equation}
\begin{split}
\label{app:Iab_def}
{\bar h}_{\1\a\b}^< (u,r,{\hat x}) &\equiv \frac{1}{8\pi r}  \int_0^{2\rho}  \dt s \int_{\frac{s}2}^{\rho}  \dt r' r' \int_0^{2\pi}  \dt \phi T_{\0\a\b} \left( u - s , r' , \frac{s(s+2r-2r')}{2 r r'} , \phi \right) , \\
{\bar h}_{\1\a\b}^> (u,r,{\hat x}) &\equiv \frac{1}{8\pi r}  \int_0^\infty  \dt s  \int_{\max(\frac{s}2,\rho)}^{\frac{s}2+r}  \dt r' r' \int_0^{2\pi}  \dt \phi T_{\0\a\b} \left( u - s , r' , \frac{s(s+2r-2r')}{2 r r'} , \phi \right) .
\end{split}
\end{equation}

In the first integral, we take $r \gg \rho$. The integration ranges for $s$ and $r'$ are both bounded and we can perform the large $r$ expansion of the integrand pointwise. It then follows that 
\begin{equation}
\begin{split}
{\bar h}_{\1\a\b}^< (u,r,{\hat x}) &= \frac{1}{8\pi r} \int_{u-2\rho}^u \!\!\!\!\!\!\!\! \dt u' \int_{D_\rho(u-u')} \!\!\!\!\!\!\!\!\!\!\!\! \dt \O_2' \frac{(u-u')^2}{\l^3} T_{\0\a\b}  \left( u' , \frac{u-u'}{\l} , {\hat x}' \right)  + \frac{B_{\a\b}(u,\rho,{\hat x})}{r^2} + \CO_\rho(r^{-3}) ,
\end{split}
\end{equation}
where $D_\rho(s) \subset S^2$ is the region defined by $\frac{s}{\rho} < \l < 2$, and
\begin{equation}
\begin{split}
B_{\a\b}(u,\rho,{\hat x}) \equiv \frac{1}{16\pi}  \int_0^{2\rho}  \dt s \int_{\frac{s}2}^{\rho}  \dt r' s(s-2r') \left( \p_\l \int_0^{2\pi}  \dt \phi T_{\0\a\b} ( u - s , r' , \l , \phi ) \right) \bigg|_{\l=\frac{s}{r'}}  . 
\end{split}
\end{equation}
$O_\rho(r^{-3})$ means that we are considering the large $r$ limit keeping $\rho$ fixed. However, we now wish to relax this condition and allow $\rho \equiv \rho(r)$. We choose this function so that at large $r$
\begin{equation}
\begin{split}
\rho(r) \to \infty, \qquad \frac{\rho(r)}{r} = o(1)  .
\end{split}
\end{equation}
Any function $\rho(r)$ satisfying these conditions can be chosen for the analysis here.\footnote{A particular example that works for our purposes is $\rho(r) = \sqrt{r L}$.} 

To proceed, we assume that for all $r > L$, the support of the stress tensor is contained in $u \in [u_- , u_+ ] \cup [ v_- - 2 r  , v_+ - 2 r ]$. The first interval corresponds to outgoing radiation and the second interval corresponds to incoming radiation. Furthermore, when $r > L$, we can write the stress-tensor expansions using the radial and logarithmic coefficient notation of Section~\ref{sec:asymp_data}:
\begin{equation}
\begin{split}
\label{app:st_exp_Cartesian}
T_{\0\a\b}(u,r,{\hat x}) &= \sum_{m=2}^\infty\sum_{n=0}^\infty \frac{(\ln\frac{r}{L})^n}{r^m}\overset{m,n}{T^{\mathrm{out}}_{\0\a\b}}(u,{\hat x}), \\
T_{\0\a\b}(v-2r,r,{\hat x}) &= \sum_{m=2}^\infty\sum_{n=0}^\infty \frac{(\ln\frac{r}{L})^n}{r^m}\overset{m,n}{T^{\mathrm{in}}_{\0\a\b}}(v,{\hat x}).
\end{split}
\end{equation}
The labels $\mathrm{out}$ and $\mathrm{in}$ distinguish the expansions at fixed $u$ and at fixed $v=u+2r$, respectively, i.e., near $\CI^+$ and $\CI^-$, respectively. For the free source $T_{\0\a\b}$ considered here, the leading $1/r^2$ coefficients have $n=0$ and hence no dependence on $\ln \frac{r}{L}$. The advanced coordinates $(v,r,{\hat x})$ obey $X^\a=(v-2r)t^\a+r q^\a({\hat x})$. $T_{\0vv}$ denotes the $vv$ component in this chart.

Using these assumptions for $\rho(r)$ and the stress tensor, we find
\begin{equation}
\begin{split}
{\bar h}_{\1\a\b}^< (u,r,{\hat x}) &= \frac{1}{8\pi r} \int_{u-2\rho}^u \dt u' \int_{D_\rho(u-u')} \dt \O_2' \frac{(u-u')^2}{\l^3} T_{\0\a\b}  \left( u' , \frac{u-u'}{\l} , {\hat x}' \right) + o(r^{-1}) . 
\end{split}
\end{equation}

We now turn to the second integral in \eqref{app:Iab_def}. Here, we take $r$ to be sufficiently large so that $r>\rho(r) > L $. The support assumption on the stress tensor then gives 
\begin{equation}
\begin{split}
\label{app:Iab_large}
{\bar h}_{\1\a\b}^> (u,r,{\hat x}) &= \frac{1}{8\pi r} \t(u-u_-) \int_{u_-}^{\min(u,u_+)} \dt u' \int_{\max(\frac{u-u'}2,\rho)}^{\frac{u-u'}2+r}  \dt r' r'  \\
&\qquad \qquad \qquad \qquad \times \int_0^{2\pi}  \dt \phi T_{\0\a\b} \left( u' , r' , \frac{(u-u')(u-u'+2r-2r')}{2 r r'} , \phi \right) \\
&\qquad + \frac{1}{8\pi r} \t( u+2 r - v_- ) \t(v_+ - u) \int_{\max(v_-,u)}^{\min(u+2r,v_+)} \dt v\int_{\max(\frac{v-u}2,\rho)}^\infty \dt r' r'  \\
&\qquad \qquad \qquad \qquad \times \int_0^{2\pi}  \dt \phi T_{\0\a\b} \left( v - 2 r' , r' , \frac{(u-v + 2 r' )(u-v+2r)}{2 r r'} , \phi \right) .
\end{split}
\end{equation}
In both terms above, $r'$ is large in the integrand and $u'$ and $v$ are finite. Consequently, we can use \eqref{app:st_exp_Cartesian} to simplify the large $r$ limit of the above. In doing so, we find that the second term above is divergent as we are integrating to $r' =\infty$. We regulate this with the infrared cutoff $R$ and find
\begin{equation}
\begin{split}
{\bar h}_{\1\a\b}^{>,R} (u,r,{\hat x}) &= \frac{1}{4r} \left[ \ln \frac{r}{\rho(r)} \int_{-\infty}^u \dt u'  \overset{2,0}{T^{\mathrm{out}}_{\0\a\b}} (u',{\hat x}) +  \ln \frac{R}{\rho(r)} \int_u^\infty \dt v  \overset{2,0}{T^{\mathrm{in}}_{\0\a\b}} (v,-{\hat x}) \right]  + o(r^{-1}) .
\end{split}
\end{equation}
The Heaviside functions in \eqref{app:Iab_large} have been absorbed into the integration limits for $u'$ and $v$. Notice that the $out$ stress tensor is evaluated at the same angular point as the field, whereas the $in$ stress tensor is evaluated at the antipodal angular point. Putting all the results derived here together, we find
\begin{equation}
\begin{split}
\label{app:habret_final_1}
{\bar h}_{\1\a\b} (u,r,{\hat x}) &= \frac{1}{r}\left[ \frac{1}{8\pi }  \int_{-\infty}^u \dt u' \int_{D_{\rho(r)}(u-u')} \dt \O_2' \frac{(u-u')^2}{\l^3} T_{\0\a\b} \left( u' , \frac{u-u'}{\l} , {\hat x}' \right) \right. \\
&\left. \qquad \quad  + \frac{1}4  \ln \frac{r}{\rho(r)}\int_{-\infty}^u \dt u'  \overset{2,0}{T^{\mathrm{out}}_{\0\a\b}} (u',{\hat x}) + \frac{1}4  \ln \frac{R}{\rho(r)} \int_u^\infty \dt v \overset{2,0}{T^{\mathrm{in}}_{\0\a\b}} (v,-{\hat x}) \right] + o(r^{-1}) . 
\end{split} 
\end{equation}
Notice that in general, we cannot simply take $\rho(r) \to \infty$ in each of the terms above separately, since they are independently divergent. However, the dependence on $\rho(r)$ cancels in the sum to the order shown above.

We now relate the Cartesian components of the stress tensor to the components in retarded and advanced coordinates. For this, we start with the following identity
\begin{equation}
\begin{split}
\label{app:T_identity_retarded}
T_{\0\a\b} &= T_{\0rr} t_\a t_\b + ( T_{\0uu} + T_{\0rr}  -  2 T_{\0ur} ) q_\a q_\b + ( T_{\0ur} - T_{\0rr} )  ( t_\a q_\b + t_\b q_\a ) \\
&\qquad - \frac{1}{r} T_{\0rA} D^A ( t_\a q_\b  + t_\b q_\a  ) + \frac{1}{r} ( T_{\0rA} - T_{\0uA} ) D^A ( q_\a  q_\b  ) + \frac{1}{r^2} T_{\0AB} D^A q_\a D^B q_\b.
\end{split}
\end{equation}
Here, all functions are evaluated at $(u,r,{\hat x})$ and $q_\a \equiv q_\a({\hat x})$. Taking a large $r$ limit on both sides, using \eqref{grav_st_large_r} and \eqref{scalar_st_large_r}, and extracting the leading term, we find
\begin{equation}
\begin{split}
\label{app:st_form_leading}
\overset{2,0}{T^\text{out}_{\0\a\b}} (u,{\hat x}) &=  q_\a({\hat x}) q_\b({\hat x}) \overset{2,0}{T_{\0uu}} (u,{\hat x}) . 
\end{split}
\end{equation}
Similarly, in advanced coordinates, the identity is
\begin{equation}
\begin{split}
\label{app:T_identity_advanced}
T_{\0\a\b} &= ( 4T_{\0vv} + 4T_{\0vr} + T_{\0rr} ) t_\a t_\b  + ( T_{\0vv} + 2T_{\0vr} + T_{\0rr} ) q_\a q_\b \\
&\qquad  - ( 2T_{\0vv} + 3T_{\0vr} + T_{\0rr} )( t_\a q_\b + t_\b q_\a )  - \frac{1}{r} ( 2T_{\0vA} + T_{\0rA} ) D^A ( t_\a q_\b + t_\b q_\a ) \\
&\qquad  + \frac{1}{r} ( T_{\0vA} + T_{\0rA} ) D^A ( q_\a q_\b )  + \frac{1}{r^2} T_{\0AB} D^A q_\a D^B q_\b .
\end{split}
\end{equation}
Here, all the functions are evaluated at $(v,r,\hat x)$, and $q_\a \equiv q_\a({\hat x})$. The identity is slightly more complicated as the relationship between Cartesian coordinates and advanced coordinates is $X^\a = ( v - 2 r ) t^\a + r q^\a({\hat x})$. Taking a large $r$ limit on both sides, we find
\begin{equation}
\begin{split}
\overset{2,0}{T^\text{in}_{\0\a\b}}(v,{\hat x}) &=  q_\a(-{\hat x}) q_\b(-{\hat x})\overset{2,0}{T_{\0vv}}(v,{\hat x})  . 
\end{split}
\end{equation}
Here, we used the property $q_\a(-{\hat x}) = 2 t_\a - q_\a({\hat x})$. From this, it follows that 
\begin{equation}
\begin{split}
\overset{2,0}{T^\text{in}_{\0\a\b}}(v,-{\hat x}) &=  q_\a({\hat x}) q_\b({\hat x})\overset{2,0}{T_{\0vv}}(v,-{\hat x})  . 
\end{split}
\end{equation}
Using this, we find
\begin{equation}
\begin{split}
\label{app:habret_final}
&{\bar h}_{\1\a\b} (u,r,{\hat x}) = \frac{1}{r} \left[ \frac{1}{8\pi} \int_{-\infty}^u \dt u' \int_{D_{\rho(r)}(u-u')} \dt \O_2' \frac{(u-u')^2}{\l^3} T_{\0\a\b} \left( u' , \frac{u-u'}{\l} , {\hat x}' \right) \right. \\
&\qquad \left. + \frac{1}4 q_\a({\hat x}) q_\b({\hat x}) \left\{  \ln \frac{r}{\rho(r)}\int_{-\infty}^u \dt u'  \overset{2,0}{T_{\0uu}}(u',{\hat x}) +  \ln \frac{R}{\rho(r)} \int_u^\infty \dt v \overset{2,0}{T_{\0vv}}(v,-{\hat x}) \right\} \right]  + o(r^{-1}) . 
\end{split} 
\end{equation}
From this, we can evaluate the metric perturbation directly,
\begin{equation}
\begin{split}
\label{app:hab_final}
&h_{\1\a\b} (u,r,{\hat x}) = \frac{1}{r} \left[ \frac{1}{8\pi} \left( \d_\a^\c \d_\b^\d - \frac{1}{2} \eta_{\a\b} \eta^{\c\d} \right) \int_{-\infty}^u \dt u' \int_{D_{\rho(r)}(u-u')} \!\!\!\!\!\!\!\!\! \dt \O_2' \frac{(u-u')^2}{\l^3} T_{\0\c\d} \left( u' , \frac{u-u'}{\l} , {\hat x}' \right) \right. \\
&\qquad \left. + \frac{1}4 q_\a({\hat x}) q_\b({\hat x}) \left\{  \ln \frac{r}{\rho(r)}\int_{-\infty}^u \dt u'  \overset{2,0}{T_{\0uu}}(u',{\hat x}) +  \ln \frac{R}{\rho(r)} \int_u^\infty \dt v \overset{2,0}{T^{\mathrm{in}}_{\0vv}}(v,-{\hat x}) \right\} \right]  + o(r^{-1}) . 
\end{split}
\end{equation}

We now use this to evaluate the components of the retarded solution in retarded coordinates. Let us start with the $uu$ component. This is
\begin{equation}
\begin{split}
h_{\1uu}(u,r,{\hat x}) &= t^\a t^\b h_{\1\a\b} (u,r,{\hat x}) \\
&= \frac{1}{r} \left[ \frac{1}4 \ln \frac{r}{L}\int_{-\infty}^u \dt u'  \overset{2,0}{T_{\0uu}}(u',{\hat x}) - \frac{1}4 \ln \frac{R}{L} \int_{-\infty}^u \dt u'  \overset{2,0}{T_{\0uu}}(u',{\hat x})  \right. \\
&\left. \qquad + \frac{1}{8\pi} \left( t^\a t^\b  +  \frac{1}{2} \eta^{\a\b} \right) \int_{-\infty}^u \dt u' \int_{D_{\rho(r)}(u-u')} \!\!\!\!\!\!\!\!\! \dt \O_2' \frac{(u-u')^2}{\l^3} T_{\0\a\b} \left( u' , \frac{u-u'}{\l} , {\hat x}' \right) \right. \\
&\left. \qquad + \frac{1}4 \ln \frac{R}{\rho(r)} \left\{ \int_{-\infty}^u \dt u'  \overset{2,0}{T_{\0uu}}(u',{\hat x})  + \int_u^\infty \dt v \overset{2,0}{T^{\mathrm{in}}_{\0vv}}(v,-{\hat x}) \right\} \right]  + o(r^{-1}) . 
\end{split}
\end{equation}
From this, we can extract
\begin{equation}
\begin{split}
\label{app:huu11}
\overset{1,1}{h_{\1uu}}(u,{\hat x}) = \frac{1}4 \int_{-\infty}^u \dt u'  \overset{2,0}{T_{\0uu}}(u',{\hat x})  ,
\end{split}
\end{equation}
and
\begin{equation}
\begin{split}
\label{app:huu10}
\overset{1,0}{h_{\1uu}}(u,{\hat x}) &=  - \frac{1}4 \ln \frac{R}{L} \int_{-\infty}^u \dt u'  \overset{2,0}{T_{\0uu}}(u',{\hat x})   \\
&\qquad + \lim_{\rho \to \infty} \left[ \frac{1}{8\pi} \left( t^\a t^\b  +  \frac{1}{2} \eta^{\a\b} \right) \int_{-\infty}^u \!\! \dt u' \int_{D_{\rho}(u-u')} \!\!\!\!\!\!\!\!\! \dt \O_2' \frac{(u-u')^2}{\l^3} T_{\0\a\b} \left( u' , \frac{u-u'}{\l} , {\hat x}' \right) \right. \\
&\left. \qquad \qquad \qquad \qquad \qquad + \frac{1}4 \ln \frac{R}{\rho} \left\{ \int_{-\infty}^u \dt u'  \overset{2,0}{T_{\0uu}}(u',{\hat x})  + \int_u^\infty \dt v \overset{2,0}{T^{\mathrm{in}}_{\0vv}}(v,-{\hat x}) \right\}  \right]  . 
\end{split}
\end{equation}

Next, we turn to $h_{\1ur}$. This is
\begin{equation}
\begin{split}
h_{\1ur}(u,r,{\hat x}) &= t^\a q^\b({\hat x}) h_{\1\a\b} (u,r,{\hat x}) \\
&= \frac{1}{r} \left[ \frac{1}{8\pi} \left(  t^\a q^\b({\hat x}) + \frac{1}{2} \eta^{\a\b} \right) \int_{-\infty}^u \dt u' \int_{S^2} \dt \O_2' \frac{(u-u')^2}{\l^3} T_{\0\a\b} \left( u' , \frac{u-u'}{\l} , {\hat x}' \right) \right] \\
&\qquad \qquad + o(r^{-1}) . 
\end{split}
\end{equation}
From this, we can read off
\begin{equation}
\begin{split}
\overset{1,0}{h_{\1ur}}(u,{\hat x}) = \frac{1}{8\pi} \left(  t^\a q^\b({\hat x}) + \frac{1}{2} \eta^{\a\b} \right) \int_{-\infty}^u \dt u' \int_{S^2} \dt \O_2' \frac{(u-u')^2}{\l^3} T_{\0\a\b} \left( u' , \frac{u-u'}{\l} , {\hat x}' \right) .
\end{split}
\end{equation}
To simplify this further, we change the integration variable from $u'$ to $r' = \frac{u-u'}{\l}$ and then to Cartesian coordinates, $\vec{X} = r' {\hat x}'$. We then find
\begin{equation}
\begin{split}
\label{app:int_prop}
&\int_{-\infty}^u \dt u' \int_{S^2} \dt \O_2' \frac{(u-u')^2}{\l^3} T_{\0\a\b} \left( u' , \frac{u-u'}{\l} , {\hat x}' \right) = \int \dt^4 X \d(u  + q({\hat x}) \cdot X   ) T_{\0\a\b}(X) . 
\end{split}
\end{equation}
Then,
\begin{equation}
\begin{split}
\label{app:hur10}
\overset{1,0}{h_{\1ur}}(u,{\hat x}) = - \frac{1}{8\pi} t^\a P_{\0\a} + \frac{1}{16\pi}  \int \dt^4 X \d(u  + q({\hat x}) \cdot X   ) \eta^{\a\b} T_{\0\a\b}(X). 
\end{split}
\end{equation}
Here, $P_\0^\a = ( E_\text{tot} , \vec{P}_\text{tot}) $ is the total free energy-momentum through a constant time slice,
\begin{equation}
\begin{split}
\label{app:Pa_def}
P_{\0\a}  \equiv - q^\b({\hat x}) \int \dt^4 X \d(u  + q({\hat x}) \cdot X   ) T_{\0\a\b}(X)  .
\end{split}
\end{equation}
This is independent of $u$ and $x^A$ due to stress tensor conservation. To see this, we note that
\begin{equation}
\begin{split}
\p_u P_{\0\a}  &= - q^\b({\hat x}) \int \dt^4 X \p_u \d( u  + q({\hat x}) \cdot X   )  T_{\0\a\b}(X) \\
&= - \int \dt^4 X \p^\b \d( u  + q({\hat x}) \cdot X ) T_{\0\a\b}(X) \\
&= \int \dt^4 X \d( u  + q({\hat x}) \cdot X ) \p^\b T_{\0\a\b}(X) \\
&= 0 ,
\end{split}
\end{equation}
and
\begin{equation}
\begin{split}
\p_A P_{\0\a} &= - \int \dt^4 X [ \p_A \d(u  + q({\hat x}) \cdot X   ) q^\b({\hat x}) T_{\0\a\b}(X) + \d(u  + q({\hat x}) \cdot X   ) \p_A q^\b({\hat x}) T_{\0\a\b}(X) ] \\
&= - \int \dt^4 X [ \p^\b \d(u  + q({\hat x}) \cdot X   ) \p_A q({\hat x}) \cdot X T_{\0\a\b}(X) +  \d(u  + q({\hat x}) \cdot X   ) \p_A q^\b({\hat x}) T_{\0\a\b}(X) ] \\
&= \int \dt^4 X \d(u  + q({\hat x}) \cdot X   ) [ \p_A q({\hat x}) \cdot X ]  \p^\b T_{\0\a\b}(X) \\
&= 0 .
\end{split}
\end{equation}
Since $P_{\0\a}$ is $u$-independent, we can take $u \to \infty$. Then, using \eqref{app:st_form_leading} and \eqref{app:int_prop}, we have an alternative form for $P_{\0\a}$,
\begin{equation}
\begin{split}
P_{\0\a} &\equiv \int_{-\infty}^\infty \dt u' \int_{S^2} \dt \O_2' q_\a({\hat x}') \overset{2,0}{T_{\0uu}} (u',{\hat x}')  . 
\end{split}
\end{equation}

Next, we turn to $h_{\1uA}$. This is
\begin{equation}
\begin{split}
h_{\1uA}(u,r,{\hat x}) &= r t^\a \p_A q^\b({\hat x}) h_{\1\a\b} (u,r,{\hat x}) \\
&= \frac{1}{8\pi} t^\a \p_A q^\b({\hat x})  \int_{-\infty}^u \dt u' \int_{S^2} \dt \O_2' \frac{(u-u')^2}{\l^3} T_{\0\a\b} \left( u' , \frac{u-u'}{\l} , {\hat x}' \right)  + o(1) . 
\end{split}
\end{equation}
Thus,
\begin{equation}
\begin{split}
\overset{0,0}{h_{\1uA}}(u,{\hat x}) &= \frac{1}{8\pi} t^\a \p_A q^\b({\hat x})  \int_{-\infty}^u \dt u' \int_{S^2} \dt \O_2' \frac{(u-u')^2}{\l^3} T_{\0\a\b} \left( u' , \frac{u-u'}{\l} , {\hat x}' \right) .
\end{split}
\end{equation}

Next, we turn to $h_{\1rr}$. This is
\begin{equation}
\begin{split}
h_{\1rr}(u,r,{\hat x}) &= q^\a({\hat x}) q^\b({\hat x}) h_{\1\a\b} (u,r,{\hat x}) = \frac{1}{r} \left[ - \frac{1}{8\pi}  q^\a({\hat x}) P_{\0\a} \right]  + o(r^{-1}) . 
\end{split}
\end{equation}
Here, we have used \eqref{app:int_prop} and \eqref{app:Pa_def}. Therefore,
\begin{equation}
\begin{split}
\label{app:hrr10}
\overset{1,0}{h_{\1rr}}({\hat x}) = - \frac{1}{8\pi}  q^\a({\hat x}) P_{\0\a} . 
\end{split}
\end{equation}
Notice that this is $u$-independent.

Next, we turn to $h_{\1rA}$. This is
\begin{equation}
\begin{split}
h_{\1rA}(u,r,{\hat x}) &= r q^\a({\hat x}) \p_A q^\b({\hat x}) h_{\1\a\b} (u,r,{\hat x})= - \frac{1}{8\pi} \p_A q^\a({\hat x}) P_{\0\a} + o(1) . 
\end{split}
\end{equation}
Therefore,
\begin{equation}
\begin{split}
\label{app:hrA00}
\overset{0,0}{h_{\1rA}}({\hat x}) = - \frac{1}{8\pi}  \p_A q^\a({\hat x}) P_{\0\a} = \p_A \overset{1,0}{h_{\1rr}}({\hat x})  . 
\end{split}
\end{equation}

Finally, we turn to $h_{\1AB}$. This is
\begin{equation}
\begin{split}
h_{\1AB}(u,r,{\hat x}) &= r^2 \p_A q^\a({\hat x}) \p_B q^\b({\hat x}) h_{\1\a\b} (u,r,{\hat x}) \\
&= r \frac{1}{8\pi} \left( \p_A q^\a({\hat x}) \p_B q^\b({\hat x})  - \frac{1}{2} \g_{AB}({\hat x}) \eta^{\a\b} \right)  \int \dt^4 X \d(u  + q({\hat x}) \cdot X   ) T_{\0\a\b}(X) \\
&\qquad + o(r) . 
\end{split}
\end{equation}
We can therefore read off
\begin{equation}
\begin{split}
\label{hab_TT}
\overset{-1,0}{h_{\1AB}}(u,{\hat x}) = \frac{1}{8\pi} \left( \p_A q^\a({\hat x}) \p_B q^\b({\hat x})  - \frac{1}{2} \g_{AB}({\hat x}) \eta^{\a\b} \right)  \int \dt^4 X \d(u  + q({\hat x}) \cdot X   ) T_{\0\a\b}(X) . 
\end{split}
\end{equation}
The trace of this field does admit a simpler expression
\begin{equation}
\begin{split}
\g^{AB}({\hat x}) \overset{-1,0}{h_{\1AB}}({\hat x}) = - \frac{1}{8\pi} [ 2 t^\a - q^\a({\hat x}) ] P_{\0\a} = ( D^2 + 1 )  \overset{1,0}{h_{\1rr}}({\hat x})  .
\end{split}
\end{equation}

This completes the asymptotic analysis at $O(\k)$.

At $O(\k^2)$, the retarded solution has the same form as the one at $O(\k)$,
\begin{equation}
\begin{split}
\label{app:h_ret_sol_appendix_1}
{\bar h}_{\2\a\b}(X)  = \frac{1}{4\pi} \int \dt^4 X' \d((X-X')^2) \t(X^0-X'^0) T_{\1\a\b}(X') .
\end{split}
\end{equation}
The stress tensor appearing on the RHS is now the one at $O(\k)$, including contributions from both the graviton and scalar. Due to the overleading fall-offs of the metric perturbation at $O(\k)$, $T_{\1\mu\nu}$ also develops overleading fall-offs compared to $T_{\0\mu\nu}$. These are
\begin{equation}
\label{app:grav_st_large_r_1}
\badat2
T_{\1uu}=O(r^{-2}),\qquad &T_{\1ur}=O(r^{-3}), \qquad T_{\1uA}=O(r^{-1}),
\\
T_{\1rr}=O(r^{-4}), \qquad & T_{\1rA}=O(r^{-2}),\qquad T_{\1AB}=O(r^0).
\eadat
\end{equation}
It can be checked that these new fall-offs do \emph{not} modify \eqref{app:st_exp_Cartesian}. It follows then that \eqref{app:habret_final_1} holds at this order as well,
\begin{equation}
\begin{split}
\label{app:habret_final_2} 
{\bar h}_{\2\a\b} (u,r,{\hat x}) &= \frac{1}{r}\left[ \frac{1}{8\pi }  \int_{-\infty}^u \dt u' \int_{D_{\rho(r)}(u-u')} \dt \O_2' \frac{(u-u')^2}{\l^3} T_{\1\a\b} \left( u' , \frac{u-u'}{\l} , {\hat x}' \right) \right. \\
&\left. \qquad \quad  + \frac{1}4  \ln \frac{r}{\rho(r)}\int_{-\infty}^u \dt u'  \overset{2,0}{T^{\mathrm{out}}_{\1\a\b}} (u',{\hat x}) + \frac{1}4  \ln \frac{R}{\rho(r)} \int_u^\infty \dt v \overset{2,0}{T^{\mathrm{in}}_{\1\a\b}} (v,-{\hat x}) \right] + o(r^{-1}) . 
\end{split} 
\end{equation}
The equation that does receive corrections at this order is \eqref{app:st_form_leading}. Using \eqref{app:T_identity_retarded} for $T_\1$ and using \eqref{app:grav_st_large_r_1}, we now find
\begin{equation}
\begin{split}
\label{app:st_form_leading_1}
\overset{2,0}{T_{\1\a\b}} (u',{\hat x}) &= q_\a({\hat x}) q_\b({\hat x}) \overset{2,0}{T_{\1uu}}(u',{\hat x}) \\
&\qquad \qquad \qquad - 2 q_{(\a}({\hat x}) D^A q_{\b)}({\hat x}) \overset{1,0}{T_{\1uA}}(u',{\hat x})  + D^A q_\a({\hat x}) D^B q_\b({\hat x})\overset{0,0}{T_{\1AB}}(u',{\hat x}) ,
\end{split}
\end{equation}
with a similar formula holding for the stress tensor on $\CI^-$.

With these corrections, we can now evaluate leading large $r$ coefficients for all the components. We immediately see that for ${\bar h}_{\2uu}$, ${\bar h}_{\2ur}$, ${\bar h}_{\2rr}$, and ${\bar h}_{\2rA}$, the extra terms in \eqref{app:st_form_leading_1} drop out, due to the contractions with $t^\a$ and $q^\a({\hat x})$. From this, we find
\begin{equation}
\begin{split}
\label{app:hrA100}
\overset{1,0}{h_{\2rr}}({\hat x}) = - \frac{1}{8\pi}  q^\a({\hat x}) P_{\1\a}  , \quad \overset{0,0}{h_{\2rA}}({\hat x}) = \p_A \overset{1,0}{h_{\2rr}}({\hat x}) , \quad \g^{AB} \overset{-1,0}{h_{\2AB}}({\hat x}) = ( D^2 + 1 ) \overset{1,0}{h_{\2rr}}({\hat x}).
\end{split}
\end{equation}
Consequently, equations \eqref{app:huu11}, \eqref{app:huu10}, \eqref{app:hur10}, and \eqref{app:hrr10} generalise simply by replacing $T_\0 \to T_\1$. The extra terms contribute extra logarithmic corrections to ${\bar h}_{\2uA}$ and ${\bar h}_{\2AB}$, which we can extract rather easily
\begin{equation}
\begin{split}
\overset{0,1}{{\bar h}_{\2uA}}(u,{\hat x})&= \frac{1}4  \int_{-\infty}^u \dt u' \overset{1,0}{T_{\1uA}}(u',{\hat x}) , \qquad \overset{-1,1}{{\bar h}_{\2AB}} (u,{\hat x}) = \frac{1}4  \int_{-\infty}^u \dt u'   \overset{0,0}{T_{\1AB}}(u',{\hat x})  . 
\end{split} 
\end{equation}
Explicitly,
\begin{equation}
\begin{split}
\overset{1,0}{T_{\1uA}}(u,{\hat x}) &= - 2 D^B \overset{1,0}{h_{\1rr}}({\hat x}) \p_u^2 \overset{-1,0}{h_{\0AB}}(u,{\hat x}) , \qquad \overset{0,0}{T_{\1AB}}(u,{\hat x}) = - 2 \overset{1,0}{h_{\1rr}}({\hat x}) \p_u^2 \overset{-1,0}{h_{\0AB}}(u,{\hat x}) . \\
\end{split}
\end{equation}
Using this, we find
\begin{equation}
\begin{split}
\overset{0,1}{h_{\2uA}}(u,{\hat x})&= - \frac{1}{2} D^B \overset{1,0}{h_{\1rr}}({\hat x}) \p_u \overset{-1,0}{h_{\0AB}}(u,{\hat x}), \qquad \overset{-1,1}{h_{\2AB}} (u,{\hat x}) = - \frac{1}{2} \overset{1,0}{h_{\1rr}}({\hat x}) \p_u \overset{-1,0}{h_{\0AB}}(u,{\hat x}) . 
\end{split} 
\end{equation}

\printbibliography

\end{document}